\documentclass[
 reprint,
 amsmath,amssymb,
 aps,
 prc,
 showpacs,
 superscriptaddress,
 groupedaddress,
 floatfix
]{revtex4-1}

\usepackage{graphicx}
\usepackage{dcolumn}
\usepackage{bm}
\usepackage{hyperref}

\begin{document}

\preprint{APS/123-QED}

\title{Measurements of \boldmath${}^{nat}Pb(p,xn){}^{201-207}Bi$, ${}^{204}Pb(p,1-4n){}^{201-204}Bi$ and ${}^{206}Pb(p,3n){}^{204}Bi$ cross-sections at astrophysical energies ($E_p \leq 30 \, MeV$) }

\author{P.V.~Guillaumon}
    \email{guillaumon@if.usp.br}
\author{I.D.~Goldman}
\author{V.R.~Vanin}
 \affiliation{Universidade de Sao Paulo Instituto de Fisica \\ Rua do Matão 1371, 05508-090 Sao Paulo, Brazil}
\author{H.~Barcellos de Oliveira}
\affiliation{Instituto de Pesquisas Energéticas e Nucleares/Comissão Nacional de Energia Nuclear\\ Av. Prof. Lineu Prestes, 2242, 05508-000 Sao Paulo, Brazil}

\date{\today}

\begin{abstract}
Cross-sections for ${}^{nat}Pb(p,xn){}^{201-207}Bi$, ${}^{204}Pb(p,1-4n){}^{201-204}Bi$ and ${}^{206}Pb(p,3n){}^{204}Bi$ reactions have been determined in the astrophysical energy range $20-30 \, MeV$. The analysis were performed by $\gamma$-spectroscopy associated with half-lives measurements. The results were compared with previously experimental data, when available, and with theoretical calculations performed using TALYS code. We report a possible new 
$\gamma$ transition from ${}^{204m}{Pb}$ and other theoretical discrepancies, probably due overestimation of the Coulomb barrier and neutron binding energy. Also reported are new determinations of the energies and intensities of the $\gamma$'s emitted in the decays of  ${}^{205,206}Bi$. We discuss possible implications for the r/rp-processes in neutrino-driven wind in supernovae and neutron stars mergers.

\end{abstract}

\maketitle

\section{Introduction}

The rp-process is not fully understood in part due lack of experimental data in the proton-rich region. There are several p-only elements like ${}^{196}Hg$, ${}^{190}Pt$, ${}^{184}Os$, and ${}^{180}W$, all near $Z \sim 83$. If an rp-process is responsible for the nucleosynthesis of these isotopes, all the others nearby may be affected too, and could act as poisons of the protons' flux. These nuclear constraints are important for astrophysical models of supernovae and neutron mergers, where this process should takes place.

If the local temperature under supernovae collapse reaches $T>3 \times 10^{9} \, K$ followed by a proton ``freeze-out'', $(p,xn)$  reactions would be important and could enrich p-elements abundances. There are discrepancies on nuclear structure and cross-sections values for neutron deficient heavy elements. Also, there are no measured cross-sections on ${}^{204}Pb(p,xn)$ reactions on the threshold, neither for ${}^{nat}Pb(p,xn){}^{207}Bi$. The same holds for a bunch of other nuclei in the lead region. 

Precise measurements are important as constraints for nuclear-structure calculations and phenomenological models, specially in a region that is not well established yet, and could help to predict properties of other p-nuclides whose cross-sections are almost unknown. Despite its astrophysical importance, there is limited data available, the majority of them performed more than 40 years ago.

The measurement of $(p,xn)$ reactions could have other applications too: the discovery of new viable paths for radio-pharmaceutical isotopes' production. For example, the important ${}^{201}Tl$ radionuclide, whose production is, in general, made by ${}^{203}Tl(p,2n){}^{201}Pb$, could also be produced by the ${}^{204}Pb(p,3n){}^{201}Pb$ route.

For ultra-low background measurements like those used in dark matter experiments, high energy cosmic rays could penetrate the shield, after loosing energy, and provoke $(p,xn)$ and $(p,\gamma)$ reactions. Although not likely probable, could be enough to increase the noise.

\section{Experimental Details}

The samples for these experiments consisted of natural Pb metal foils of $<0.5 \, mm$ thickness and $10 \, mm$ diameter ($ > 99.9 \%$ pure). Samples were irradiated at IPEN/CNEN-SP cyclotron facility with protons from $18$ to $30 \, MeV$ and beam current of a few $\mu A$. Since this cyclotron operates only with liquid targets, an adapter to solid ones was made and attached to the target holder.

Gamma spectroscopy with an HPGe detector was performed immediately after irradiation. The samples were measured continuously and the spectra were saved every $30 \, min$, allowing us to follow half-lives from $57.5 \, min$ (${}^{201m}Bi$) to $31.55 \, y$ (${}^{207}Bi$). The detector resolution at 1332 keV was 2 keV and was coupled to a digital signal processor-based data acquisition system. Energy calibration was done with a second degree polynomial fit based on a weighted non-linear principal component analysis, \cite{Guillaumon_2019}. The peaks were fitted to Gaussian shapes with tail and linear background using IDEFIX software, \cite{Idefix}. Self-absorption of protons and $\gamma$ were estimated to be minimal and were not taken into account. The detector was shielded with $\sim 10 \, cm$ lead, leading to a contribution for the peaks of less than $0.02 \, Bq$.

End-of-bombardment activities ranged from 20 to 40 kBq with a dead-time up to $70 \%$ at $20 \, cm$ from the detector and were corrected by a $\Delta_{real}/\Delta_{live}$ term, where $\Delta_{live}$ is the live time and $\Delta_{real}$, the real one. The cross-section will be given by

\begin{equation}
    \sigma = \frac{\mathcal{A} e^{\lambda \Delta t}}{\mathcal{N}_0 \Phi \varepsilon I_\gamma \left( 1 - e^{-\lambda t_{irr}} \right)} \frac{\Delta t}{\Delta t_{live}},
\end{equation}

\noindent where $\mathcal{A}$ is the activity of the isotope, $\lambda$ the decay constant, $\mathcal{N}_0$ the number of isotopes before irradiation, $\varepsilon$ the total efficiency, $I_\gamma$ the intensity of the $\gamma$ rays, $t_{irr}$ the time interval between irradiation and measurement, $\Delta t$ the real time of measurement, and $\Delta t_{live}$ the live time.

\section{Results}

\begin{figure}[ht] 
\includegraphics[width=\textwidth]{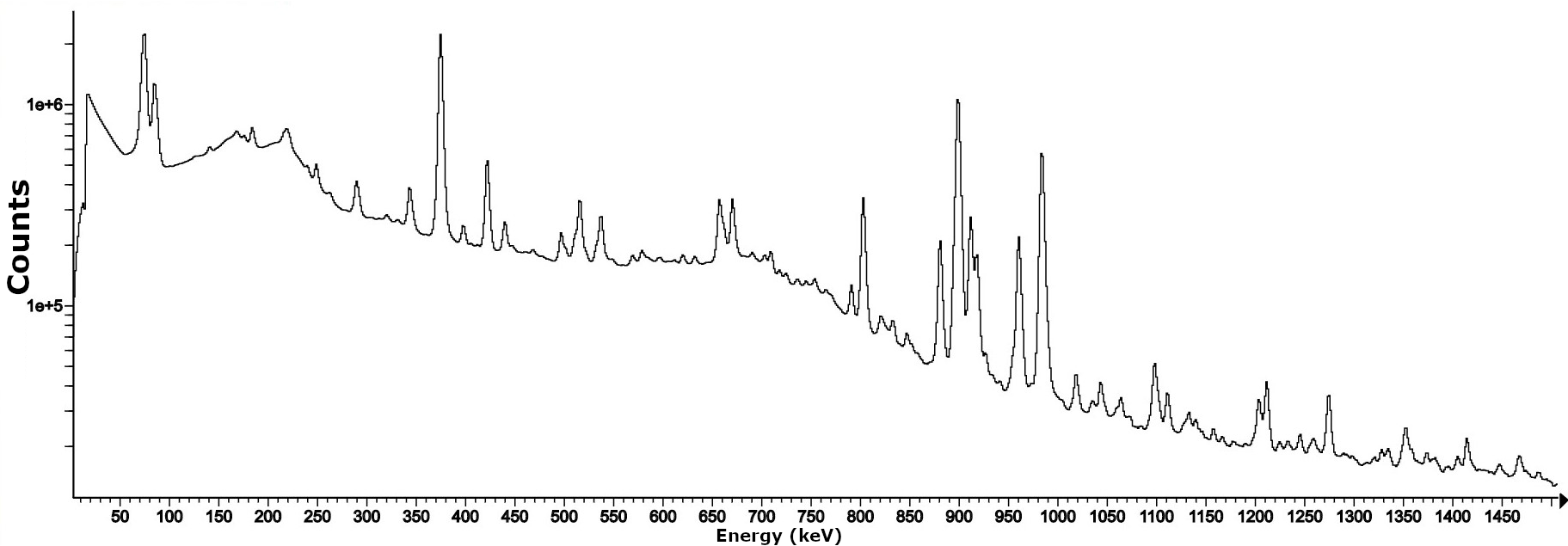}
\caption{$\gamma$ spectrum from 0 - 1.5 MeV for the irradiation with protons of 30 MeV in a a foil of natural lead. Data accumulated for 30 min and after 20 min of the irradiation.}
\label{fig:fig1}
\end{figure}

\begin{figure}[ht] 
\includegraphics[width=\textwidth]{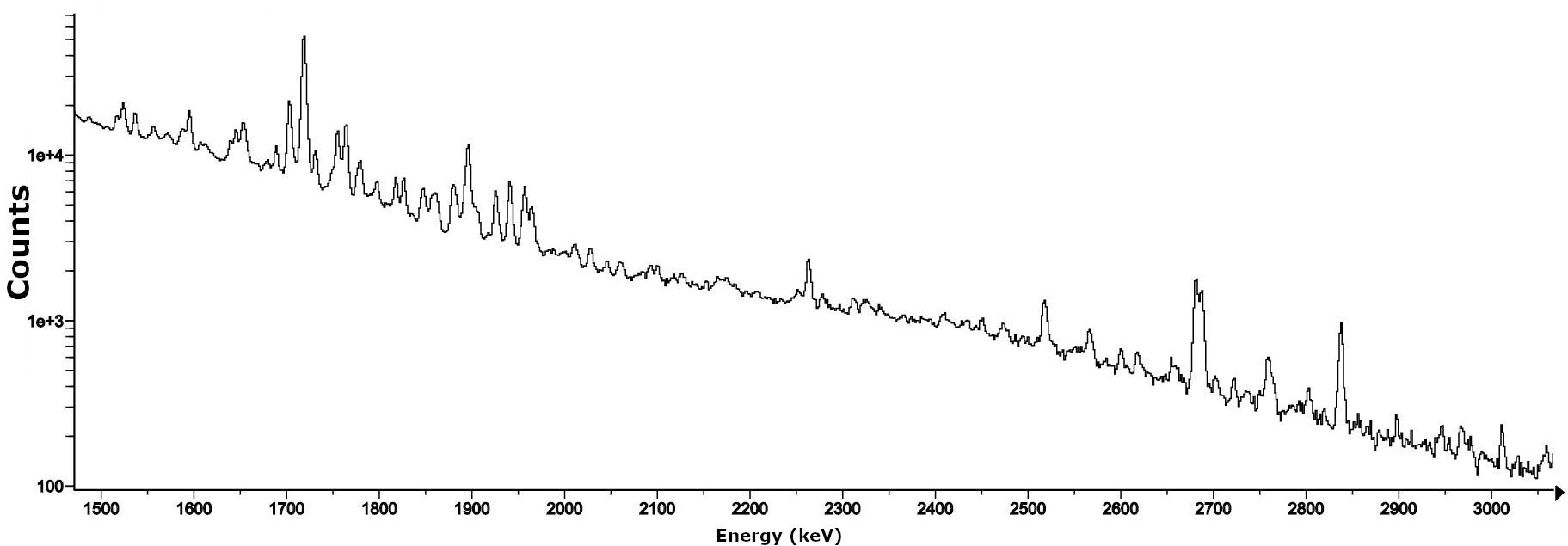}
\caption{$\gamma$ spectrum from 1.5 - 3.0 MeV for the irradiation with protons of 30 MeV in a a foil of natural lead. Data accumulated for 30 min and after 20 min of the irradiation.}
\label{fig:fig2}
\end{figure}

Figures \ref{fig:fig1} and \ref{fig:fig2} show the $\gamma$ spectra from the irradiation of ${}^{nat}Pb$ with protons of $30 \, MeV$ accumulated for about $20 \, min$ after irradiation with a live time of $30 \, min$. No appreciable impurities were observed and could be neglected. Several ${}^{201-207}Bi$ lines lies close to each other with an energy difference of less than $0.5 \, keV$. In principle, we could easily separate them by a careful measurement of the half-lives, since they range from less than one hour to a few ten years. The only exception are the lines of  ${}^{203}Bi$ ($11.76 \, h$) with ${}^{204}Bi$ ($11.22 \, h$), like the $421.8 \, keV$ line ($0.39 \%$) of the first one, and the $421.61 \, keV$ line ($1.14 \%$) of the second. To calculate reactions cross-sections it is preferably to use less intense pure lines than mixed intense ones. We used this criteria, whenever possible. The ${}^{201}Bi$ was the only exception, where there were no pure lines.

Detector efficiency and counting statistics were the dominant contribution to the uncertainty, of less than $10 \%$ together. Proton flux, sample mass, energy resolution, half-life and $\beta$ decay branching contribute with less than $5\%$ of the uncertainty.

\onecolumngrid

\twocolumngrid

\onecolumngrid

\twocolumngrid

We analyzed by gamma spectroscopy all lines with intensity greater than $1 \%$, measuring the half-life of each one in order to select the pure ones. 
For this analysis, we choose the spectrum produced by $18 \, MeV$ protons, since it is ``cleaner''.
Tables \ref{tab:i_206bi} and \ref{tab:i_205bi} summarizes the $\gamma$ intensities for ${}^{205,206}Bi$ EC decay deduced from our work in comparison with the previous works of \citet{Manthuruthil1972} and \citet{Hamilton1972} used as reference in Nuclear Data Sheets (ENDS), \cite{Kondev2004_205,Kondev2008_206}. Overall agreement between the previous results were obtained, although ours are generally more precise. The 895.12 keV ($15.67 \%$) from ${}^{206}Bi$ are overlapped by 894.56 keV ($0.622 \%$) from  ${}^{205}Bi$. We  subtracted this contribution from the peak area using the previous works in ENDS. The discrepancy in our results in comparison 
 with that of \citet{Manthuruthil1972} may be due subestimation of $\gamma$-intensity from the decay of ${}^{205}Bi$. Other cases of overlapping  are the 497.06 keV ($15.33 \%$; ${}^{206}Bi$) with the 498.87 keV ($0.040 \%$; ${}^{205}Bi$) and 498.40 keV ($0.093 \%$; ${}^{205}Bi$). In the cases of $\gamma$ intensities calculated for ${}^{205}Bi$ we observed an overlap in the 1903.45 keV ($2.47 \%$; ${}^{205}Bi$) with the 1903.56 keV ($0.349 \%$) from ${}^{206}Bi$. In all these cases the contribution of the ``poison'' gammas are much smaller and were corrected. On the other hand, the 570.60 keV ($4.34 \%$) from ${}^{205}Bi$ are overlapped by the 569.698 keV ($97.75 \%$) from ${}^{207}Bi$ (31.55 y). The uncertainty due this correction is big enough, since ${}^{207}Bi$ is largely produced from ${}^{nat}Pb(p,xn)$ reactions at $18 \, MeV$. For this reason we choose not to estimate this $\gamma$ intensity.
 
 We should also note the case of 987 keV. Although it is an overlapping of two $\gamma$'s from ${}^{205}Bi$, almost all contribution comes from the 987.66 ($16.1 \%$). We could neglect that from 987.49 keV, since it has an intensity of $0.09 \%$. The same happens with the 284 keV, since 284.15 keV ($1.69 \%$) is much more intense than that from 284.26 keV ($0.031 \%$).
 
\begin{table}
\caption{Gamma rays emitted in the decays of ${}^{206}Bi$.}
\label{tab:i_206bi}
\begin{tabular}{|c|c|c|}
\hline
$E_\gamma$ (keV) & I (This Work) & I (Previous Work) \footnote{\citet{Manthuruthil1972}; normalized.} \\ \hline
1878.65 & 0.01345 (28) & 0.0203 (5) \\
1718.70 & 0.2413 (10) & 0.322 (7) \\
1595.27 & 0.0410 (12) & 0.0507 (11) \\
1405.01 & 0.0129 (10) & 0.015 (4) \\
1098.26 & 0.1335 (10) & 0.1365 (28) \\
1018.63 & 0.0786 (8) & 0.0767 (16) \\
895.12 & 0.2044 (12) \footnote{894.56 (5) keV ($0.622 (12) \% $, absolute intensity) contribution from ${}^{205}Bi$ EC decay was subtracted,  \citet{Hamilton1972}.} & 0.158 (3) \\
881.01 & 0.6859 (19) & 0.669 (14) \\
803.10 & 1.0000 (24) & 1.000 (20) \\
657.16 & 0.0188 (8) & 0.0193 (5) \\
632.25 & 0.0427 (7) & 0.0452 (10) \\
620.48 & 0.0531 (7) & 0.0582 (12) \\
537.45 & 0.2679 (10) & 0.308 (6) \\
516.18 & 0.3629 (12) & 0.412 (8) \\
497.06 & 0.1323 (11) \footnote{498.87 (20) ($0.040 (25) \% $, absolute intensity) and 498.40 (15) keV ($0.093 (16) \% $, absolute intensity) contributions from ${}^{205}Bi$ EC decay were subtracted,  \citet{Hamilton1972}.} & 0.155 (3) \\
398.00 & 0.0873 (7) & 0.1086 (22) \\
343.51 & 0.1942 (8) & 0.237 (5) \\
262.70 & 0.0449 (5) & 0.0305 (7) \\
183.977 & 0.1370 (7) & 0.160 (4) \\ \hline
\end{tabular} 
\end{table}

\begin{table}[!htbp]
\caption{Gamma rays emitted in the decays of ${}^{205}Bi$.}
\label{tab:i_205bi}
 \begin{tabular}{|c|c|c|}
\hline
$E_\gamma$ (keV) & I (This Work) & I (Previous Work) \footnote{\citet{Hamilton1972}; normalized.} \\ \hline
1903.45 & 0.0668 (3) \footnote{1903.56 (10) keV ($0.349 (15) \% $, absolute intensity) contribution from ${}^{206}Bi$ EC decay was subtracted,  \citet{Manthuruthil1972}.} & 0.0760 (22) \\
1861.70 & 0.1772 (5) & 0.190 (6) \\
1775.80 & 0.1194 (4) & 0.123 (4) \\
1764.30 & 1.0000 (16) & 1.00 (3) \\
1619.10 & 0.01280 (27) & 0.0113 (5) \\
1614.30 & 0.0771 (5) & 0.0702 (22) \\
1551.00 & 0.0308 (6) & 0.0298 (11) \\
1548.65 & 0.0104 (4) & 0.0086 (5) \\
1351.52 & 0.0438 (5) & 0.0326 (12) \\
1190.03 & 0.0931 (5) & 0.0695 (26) \\
1043.75 & 0.3139 (8) & 0.231 (6) \\
987.66+987.49 \footnote{Main contribution comes from 987.66 keV ($16.1 (3) \%$, absolute intensity); 987.49 keV represents $0.09 (3) \%$ (absolute intensity), according to previous work.} & 0.6929 (12) & 0.498 (16) \\
910.90 & 0.0710 (4) & 0.0505 (16) \\
761.35 & 0.0272 (3) & 0.0209 (10) \\
759.10 & 0.0486 (4) & 0.0320 (17) \\
703.45 & 1.2609 (18) & 0.9569 \footnote{No uncertainty estimated by this work.} \\
579.80 & 0.2088 (5) & 0.167 (5) \\
576.30 & 0.00645 (26) & 0.00579 (25) \\
573.85 & 0.02380 (29) & 0.0191 (6) \\
549.84 & 0.1057 (4) & 0.0908 (25) \\
284.26+284.15 \footnote{Main contribution comes from 284.15 keV ($1.69 (3) \%$, absolute intensity); 284.26 keV represents $0.031 (9) \%$ (absolute intensity), according to previous work.} & 0.05626 (26) & 0.0530 (15) \\
260.50 & 0.03652 (21)  & 0.0335 (12) \\ \hline
\end{tabular}
\end{table}

\begin{table}[!htbp]
	\caption{Cross-section values $(mb)$ for proton energy in the lab frame for the reactions $ {}^{nat}{Pb}(p,xn){}^{204-207}{Bi}$.}
	\label{tab:xs-pb-nat}
	\begin{tabular}{|c|c|c|c|c|}
		\hline
		Energy (MeV) & \( {}^{207}{Bi} \) & \( {}^{206}{Bi} \) & \( {}^{205}{Bi} \) & \( {}^{204}{Bi} \) \\ \hline
		30.0 & 116.8 (19) & 459 (5) & 220 (6) & 152 (7) \\ 
		29.0 & 280.3 (25) & 755 (11) & 361 (16) & 293 (42) \\ 
		28.0 & 242 (4) & 665 (7) & 299 (12) & 202 (10) \\ 
		27.0 & - &  585 (6) & 372 (11) & 127 (6)  \\ 
		26.0 & 464 (8) & 457 (5) & 353 (11) & 56.4 (29)  \\ 
		25.0 & 859 (14) & 365 (3) & 356 (10) & 8.7 (4) \\ 
		24.0 & 799 (14) & 278.3 (28) & 296 (9) & 1.24 (7) \\ 
		23.0 & 688 (6) & 270.0 (27) & 283 (8) & 1.35 (7)  \\ 
		22.0 & 618 (6) & 238.7 (24) & 245 (7) & 1.79 (9) \\ 
		21.0 & 587 (5) & 239.4 (24) & 236 (7) & 2.49 (13)  \\ 
		20.0 & 628 (6) & 250.7 (25) & 215 (6) & 2.81 (15) \\ \hline
	\end{tabular}
\end{table}

\begin{table}[!htbp]
	\caption{Cross-section values $(mb)$ for proton energy in the lab frame for the reactions $ {}^{nat}{Pb}(p,xn){}^{201-203}{Bi}$.}
	\label{tab:xs-pb-nat-2}
	\begin{tabular}{|c|c|c|c|}
		\hline
		Energy  (MeV) & \( {}^{203}{Bi} \) & \( {}^{202}{Bi} \)  & \( {}^{201}{Bi} \) \\ \hline
		30.0 &  4.53 (17) & 4.89 (8) & 0.349 (18) \\ 
		29.0 &  8.9 (7) & 10.72 (19) & - \\ 
		28.0 &  10.0 (5) & 5.25 (9) & - \\ 
		27.0 &  12.2 (5) & 2.38 (4) & - \\ 
		26.0 & 13.2 (5) & 0.510 (9) & - \\ 
		25.0 & 14.1 (5) & - & - \\ 
		24.0 & 12.8 (5) & - & - \\ 
		23.0 & 12.3 (5) & - & - \\ 
		22.0 & 11.0 (4) & - & - \\ 
		21.0 & 8.5 (4) & - & - \\ 
		20.0 & 7.5 (3) & - & -
  \\ \hline
	\end{tabular}
\end{table}

\begin{figure}[h] 
\includegraphics[width=0.5\textwidth]{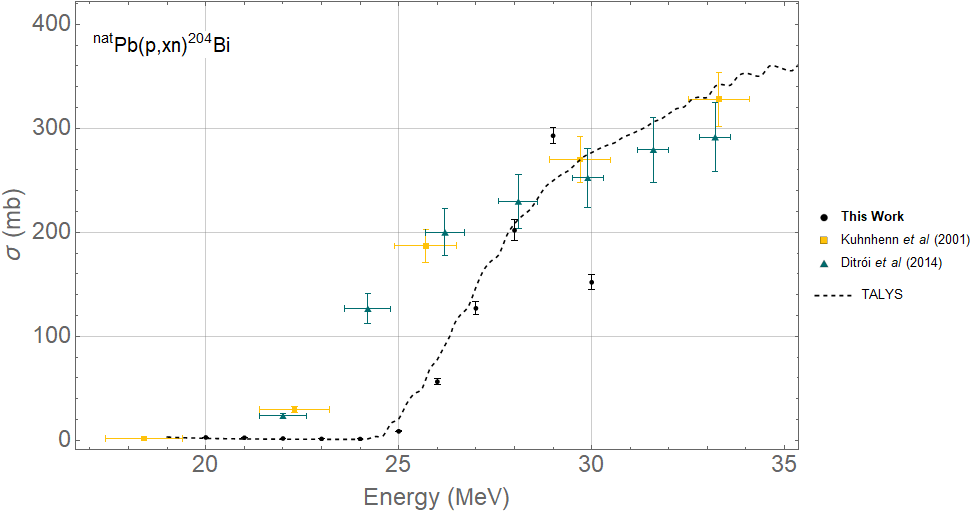}
\caption{Independent cross-sections for ${}^{nat}Pb(p,xn){}^{204}Bi$ reactions, compared with earlier experimental data together with theoretical calculations from TALYS 1.9 default code. The experimental data are taken from Refs. \cite{Kuhnhenn_2001,Ditroi_2014}. }
\label{fig:bi204}
\end{figure}

Tables \ref{tab:xs-pb-nat} and \ref{tab:xs-pb-nat-2} summarizes the resulting cross-sections obtained in our work for the reaction ${}^{nat}Pb(p,xn)$. 
In Table \ref{tab:xs-pb-206} we summarize the cross-sections results for ${}^{206}Pb(p,3n){}^{204}Bi$ obtained in this work. In Table \ref{tab:xs-pb-204} we summarize the cross sections results for ${}^{204}Pb(p,xn)$ reactions. Since the neutron separation energy of the ${}^{205}Bi$ is $8.490 \, MeV$, only ${}^{204}Pb(p,n){}^{204}Bi$ reaction occurs for protons up to 24 MeV. As we can observe in Figure \ref{fig:bi204}, after 25 MeV, the cross-section has a huge increase due the contribution of ${}^{204}Pb(p,3n){}^{204}Bi$ reactions. Until 27 MeV, ${}^{204}Pb(p,n)$ still contributes. We can perform an extrapolation to subtract this reaction from the total one. After that, all the cross-section is due ${}^{206}Pb(p,3n)$ reactions. In the case of  ${}^{201,202}Bi$, only ${}^{204}Pb(p,xn)$ reactions occurs below 30 MeV. 

\begin{figure}[h] 
\includegraphics[width=0.5\textwidth]{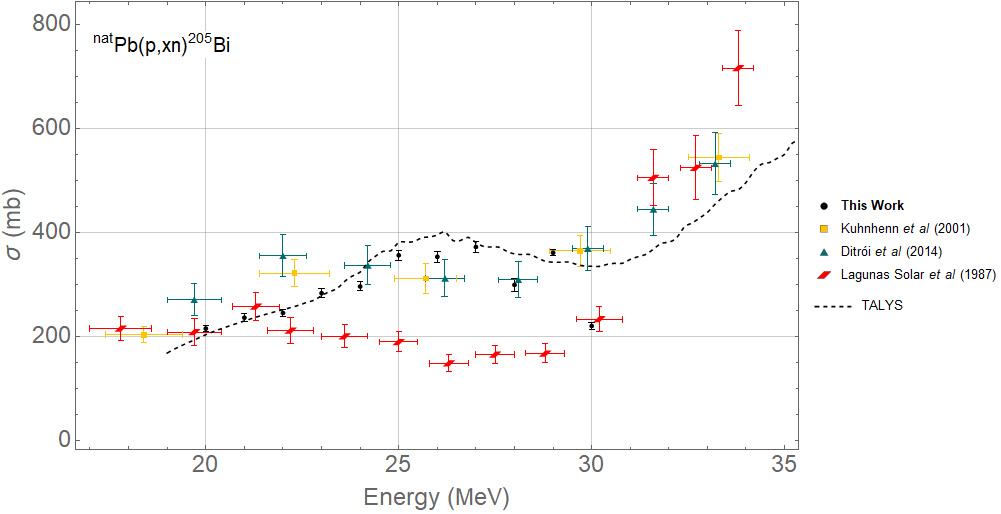}
\caption{Independent cross-sections for ${}^{nat}Pb(p,xn){}^{205}Bi$ reactions, compared with earlier experimental data together with theoretical calculations from TALYS 1.9 default code. The experimental data are taken from Refs. \cite{Kuhnhenn_2001,Ditroi_2014,Lagunas-Solar_1987}. }
\label{fig:bi205}
\end{figure}

\begin{figure}[ht] 
\includegraphics[width=0.5\textwidth]{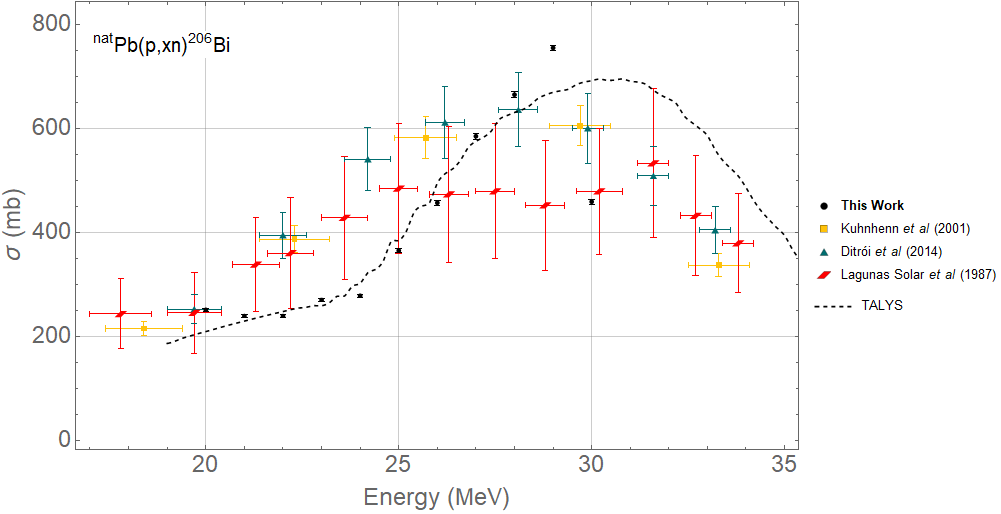}
\caption{Independent cross-sections for ${}^{nat}Pb(p,xn){}^{206}Bi$ reactions, compared with earlier experimental data together with theoretical calculations from TALYS 1.9 default code. The experimental data are taken from Refs. \cite{Kuhnhenn_2001,Ditroi_2014,Lagunas-Solar_1987}. }
\label{fig:bi206}
\end{figure}

\begin{figure}[ht] 
\includegraphics[width=0.5\textwidth]{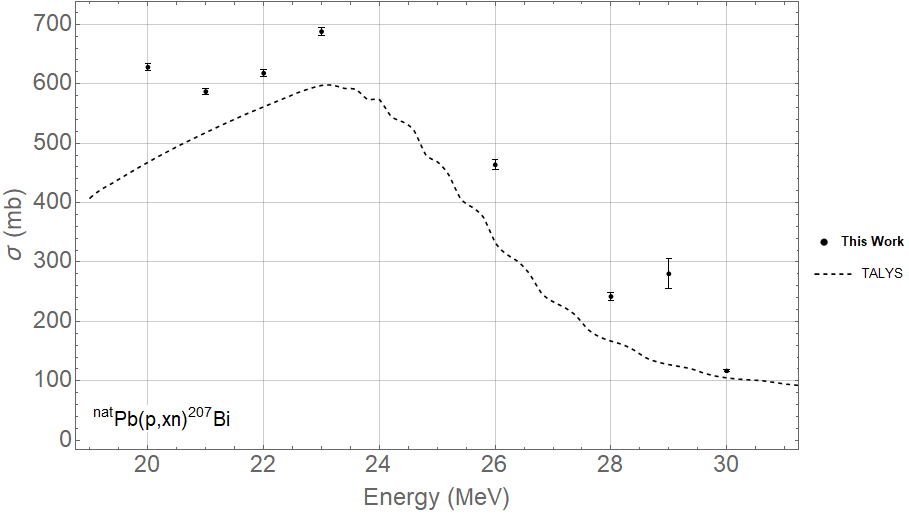}
\caption{Independent cross-sections for ${}^{nat}Pb(p,xn){}^{207}Bi$ reactions, compared with theoretical calculations from TALYS 1.9 default code. There is no previous experimental data for this reaction. }
\label{fig:bi207}
\end{figure}

\begin{figure}[ht] 
\includegraphics[width=0.5\textwidth]{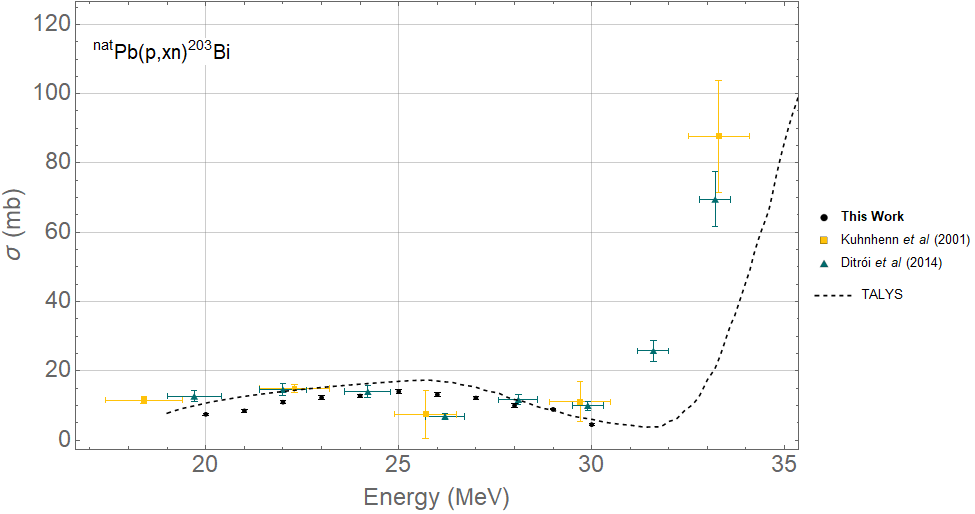}
\caption{Independent cross-sections for ${}^{nat}Pb(p,xn){}^{203}Bi$ reactions, compared with earlier experimental data together with theoretical calculations from TALYS 1.9 default code. The experimental data are taken from Refs. \cite{Kuhnhenn_2001,Ditroi_2014}. }
\label{fig:bi203}
\end{figure}

\begin{figure}[ht] 
\includegraphics[width=0.5\textwidth]{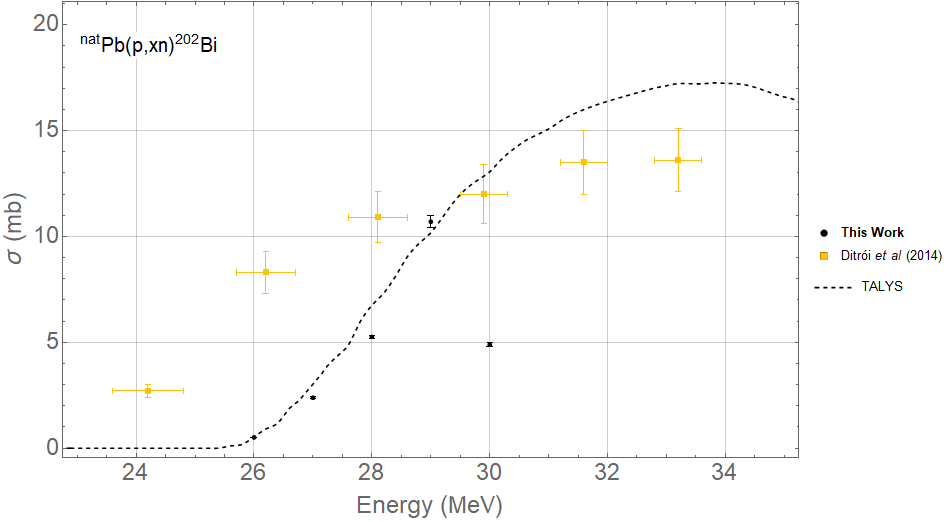}
\caption{Independent cross-sections for ${}^{nat}Pb(p,xn){}^{202}Bi$ reactions, compared with earlier experimental data together with theoretical calculations from TALYS 1.9 default code. The experimental data are taken from Ref. \cite{Kuhnhenn_2001}. }
\label{fig:bi202}
\end{figure}

\begin{figure}[ht] 
\includegraphics[width=0.5\textwidth]{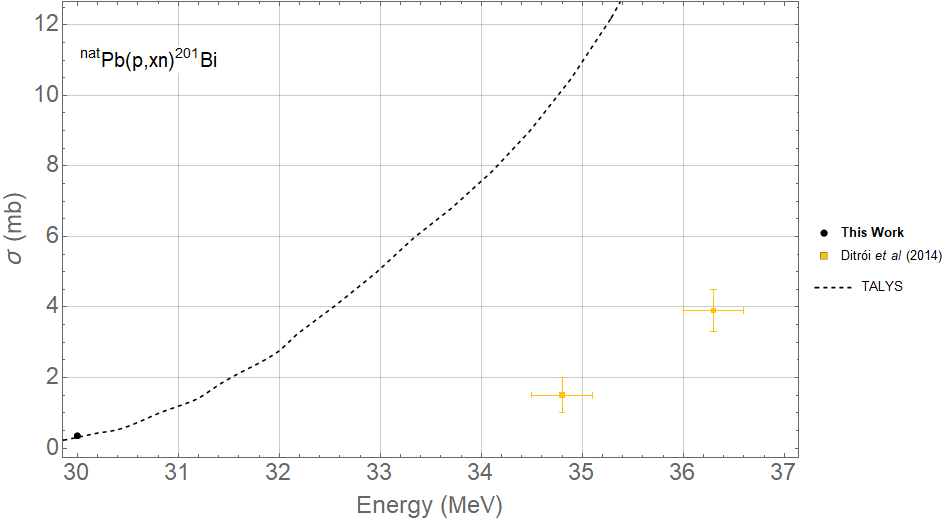}
\caption{Independent cross-sections for ${}^{nat}Pb(p,xn){}^{201}Bi$ reactions, compared with earlier experimental data together with theoretical calculations from TALYS 1.9 default code. The experimental data are taken from Ref. \cite{Kuhnhenn_2001}. }
\label{fig:bi201}
\end{figure}

In Figures \ref{fig:bi204}, \ref{fig:bi205}, \ref{fig:bi206}, \ref{fig:bi207}, \ref{fig:bi203}, \ref{fig:bi202} and  \ref{fig:bi201} we can observe the results for ${}^{nat}Pb(p,xn)$ obtained in this work in comparison with Talys 1.9 code simulations, \cite{Koning_2012}, and previous works. While the previous ones have uncertainties of $\sim 10 \%$, we increased the accuracy of the cross-sections by a factor $>5$ (uncertainties $\sim 1-5 \% $) due long measurements and half-live curves fit. In the case of ${}^{nat}Pb(p,xn){}^{207}Bi$, we have done the first measurement below 30 MeV for this reaction. 
Our results also reproduce the main features of the theoretical curve calculated by TALYS nuclear code. With exception of the ${}^{nat}Pb(p,4n){}^{201}Bi$ TALYS curve, all the others  were translated right by 4 MeV. Due the big uncertainties, previous works were not able to verify fine discrepancies in theoretical curves. We will discuss more about this later. 

\begin{figure}[ht] 
\includegraphics[width=0.5\textwidth]{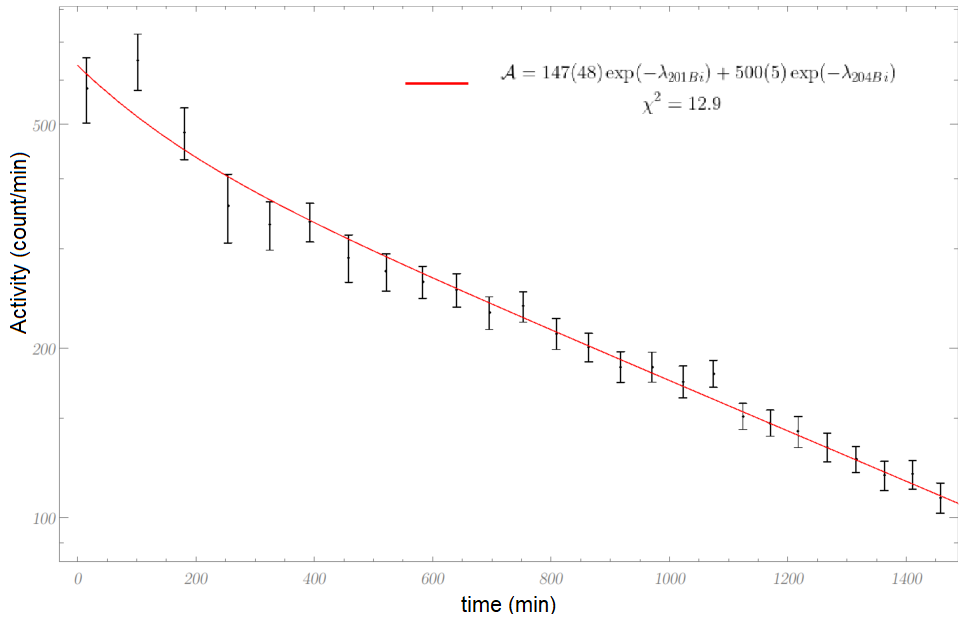}
\caption{Activity decay curve (counts/min)  for $E_\gamma = 1650.9 \, \, keV$ from ${}^{201}Bi$ ($6.3 \%$, 103 min) with $E_\gamma = 1652.10 \, \, keV$ from ${}^{204}Bi$ ($0.56 \%$, 11.22 h), obtained from irradiation of natural lead with protons of 30 MeV.}
\label{fig:bi201-1650}
\end{figure}

\begin{figure}[ht] 
\includegraphics[width=0.5\textwidth]{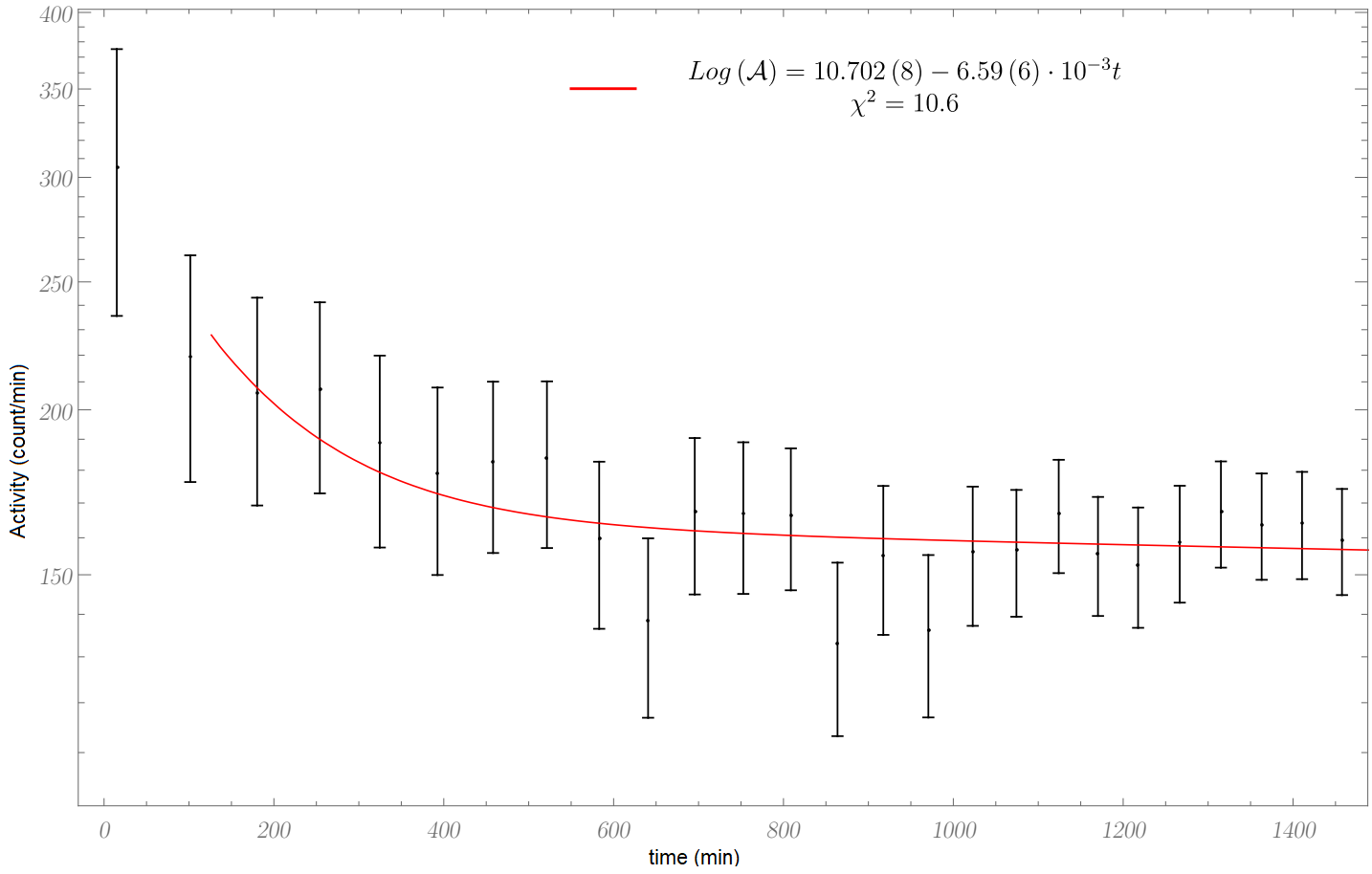}
\caption{Activity decay curve (counts/min) for $E_\gamma = 1014.1 \, \, keV$ from ${}^{201}Bi$ ($11.6 \%$, 103 min) with $E_\gamma = 1014.3 \, \, keV$ from ${}^{205}Bi$ ($0.941 \%$, 15.31 d), obtained from irradiation of natural lead with protons of 30 MeV.}
\label{fig:bi201-1014}
\end{figure}

\begin{figure}[ht] 
\includegraphics[width=0.5\textwidth]{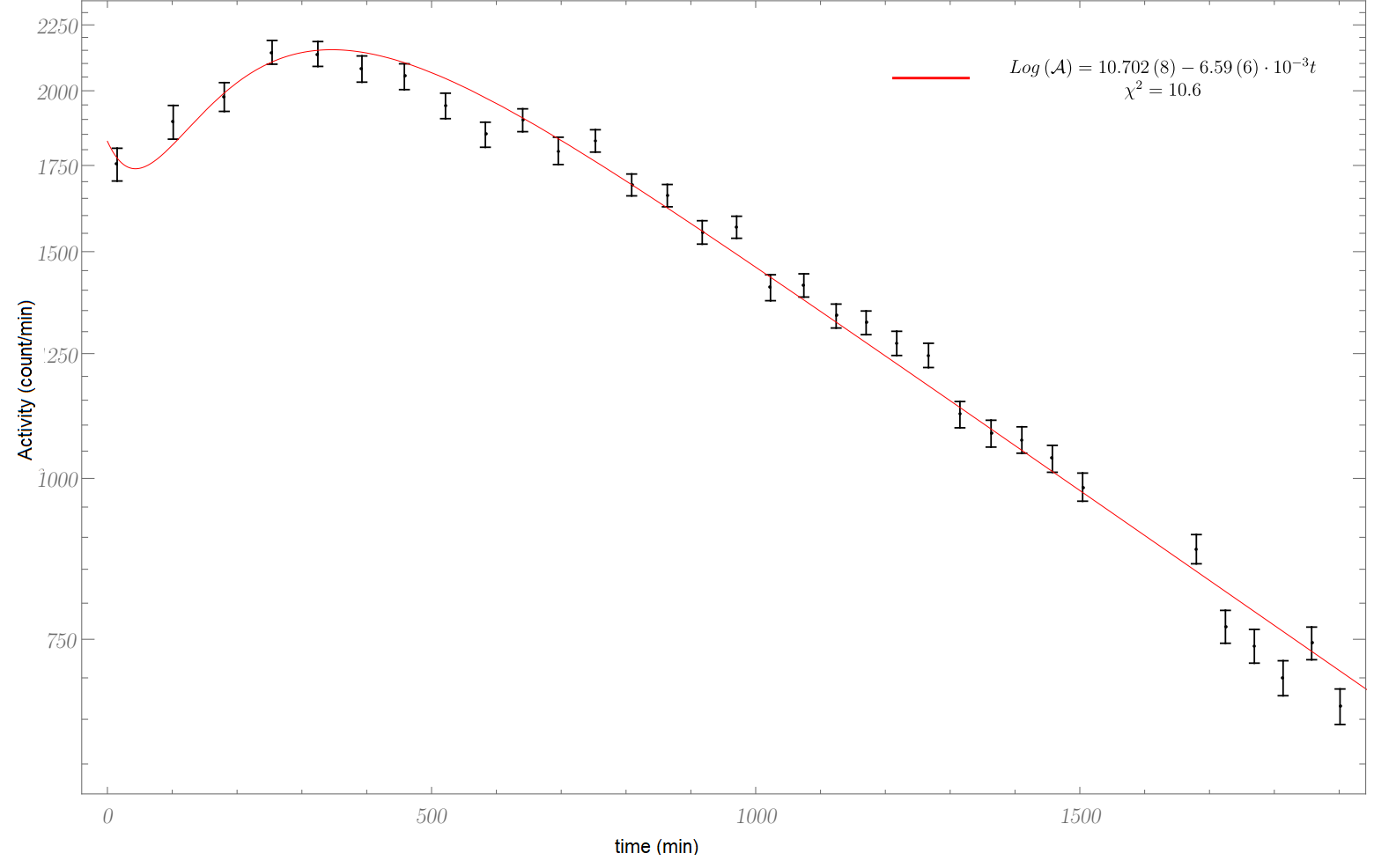}
\caption{Activity decay curve (counts/min)  for $E_\gamma = 1014 \, \, keV$ from ${}^{201}Bi$ ($1.9 \%$, 103 min)  ${}^{201m}Bi$ ($100 \%$, 57.5 min), ${}^{201m}Bi$ ($ 8.6 \%$, 11.76 h), and ${}^{204}Bi$ ($ 0.96 \%$, 11.22 h)  obtained from irradiation of natural lead with protons of 30 MeV. We could not estimate the percentage of ${}^{201m}Bi$ that is populated in this reaction.}
\label{fig:bi201-847}
\end{figure}

The first measurement of the cross-section of ${}^{204}Pb(p,4n){}^{201}Bi$ reaction in the threshold energy is of special interest for nuclear reactions models. It has two intense high energy $\gamma$'s, the 1650.9 keV ($6.3  \%$) and the 1014.1 keV ($11.6  \%$). Since ${}^{204}Pb$ has an isotopic abundance of only $1.4  \%$ in comparison to $24.1 \%$ for the ${}^{206}Pb$, and the cross-section for production of ${}^{204}Bi$ (11.22 h) on natural lead are about 400 times bigger than for the production of ${}^{201}Bi$ (103 min), the 1652.10 keV ($24.1 \%$) will lie in the same peak. This is due to the large width to have enough statistics of 1650.9 keV. In Figure \ref{fig:bi201-1650} we can see the fit of this peak assuming these two $\gamma$'s. Statistically we can assert the presence of ${}^{201}Bi$. The same holds for the 1014.1 keV, whose peak is merged with the 1014.30 keV of ${}^{205}Bi$($I_\gamma = 0.914 \%$, 15.31 d), as can be seen in Figure \ref{fig:bi201-1014}. The 847.7 keV is an emblematic one. Its peak is a sum of four  $\gamma$'s, from ${}^{201}Bi$ ($I_\gamma = 1.9\% $, 103 min), ${}^{201m}Bi$ ($I_\gamma = 100 \%$, 57.5 min), ${}^{203}Bi$ ($I_\gamma = 8.6 \% $, 11.76 h) and ${}^{204}Bi$ ($I_\gamma = 0.96 \%$, 11.22h). As we can see in Figure \ref{fig:bi201-847}, we can statistically assert the presence of ${}^{201}Bi$. We did not use the 847 keV peak to calculate the cross-section for ${}^{204}Pb(p,4n){}^{201}Bi$ for two main reasons: we did not know how much of the metastate was populated at 30 MeV and the uncertainty in this separation is big enough to have a reliable value for the cross-section.

\begin{figure}[ht] 
\includegraphics[width=0.5\textwidth]{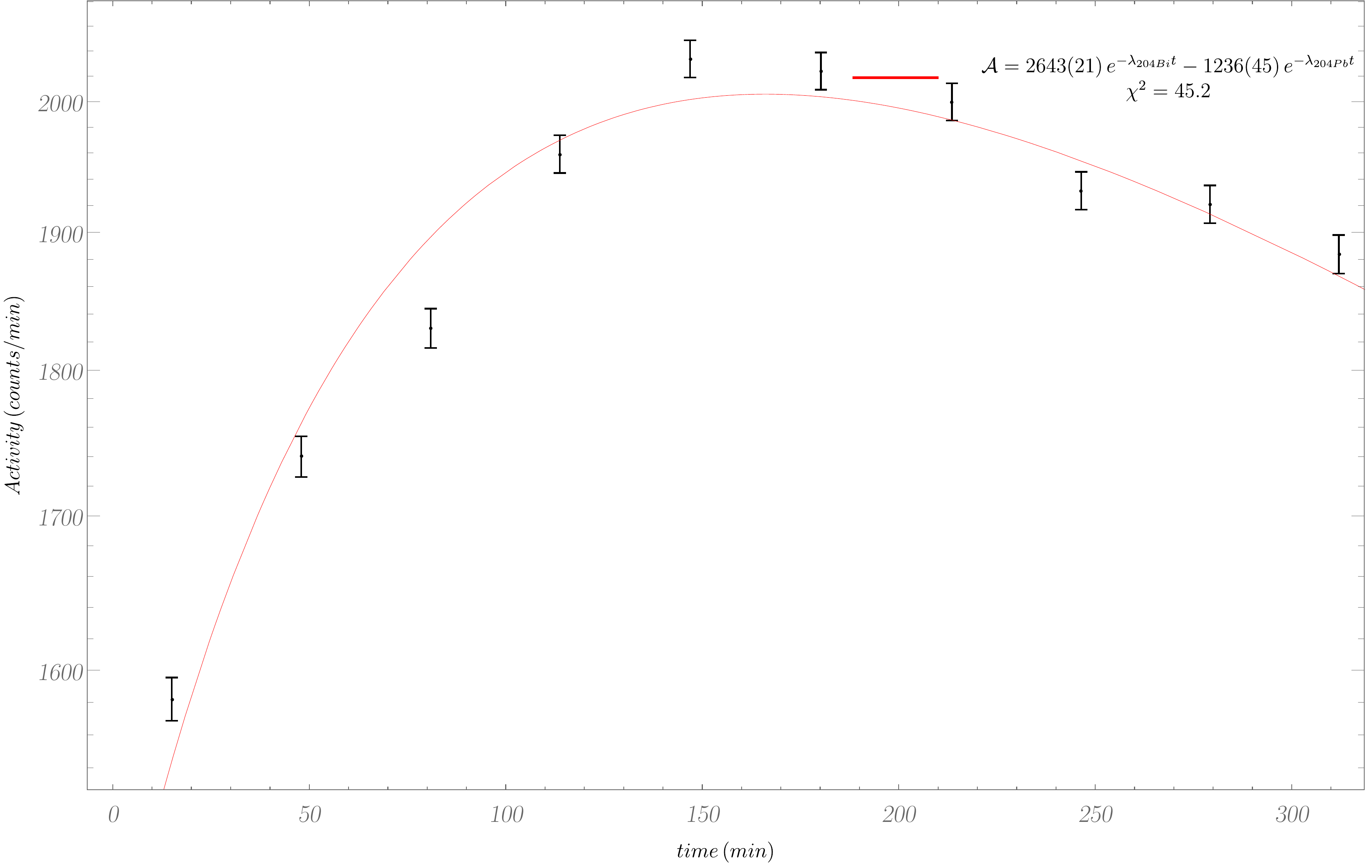}
\caption{Activity decay curve (counts/min)  for $E_\gamma = 911.96 \, \, keV$ ($11.2 \%$) and $911.74 \, \, keV$ ($13.6 \%$) from ${}^{204}Bi$. The last one comes from the $\gamma$-transition of ${}^{204m}Pb$ ($91.5 \%$, 66.93 min), as can be seen from our fit. This delayed $\gamma$ was not completely reported at ENDS.}
\label{fig:Bi204-911}
\end{figure}

\begin{figure}[ht] 
\includegraphics[width=0.5\textwidth]{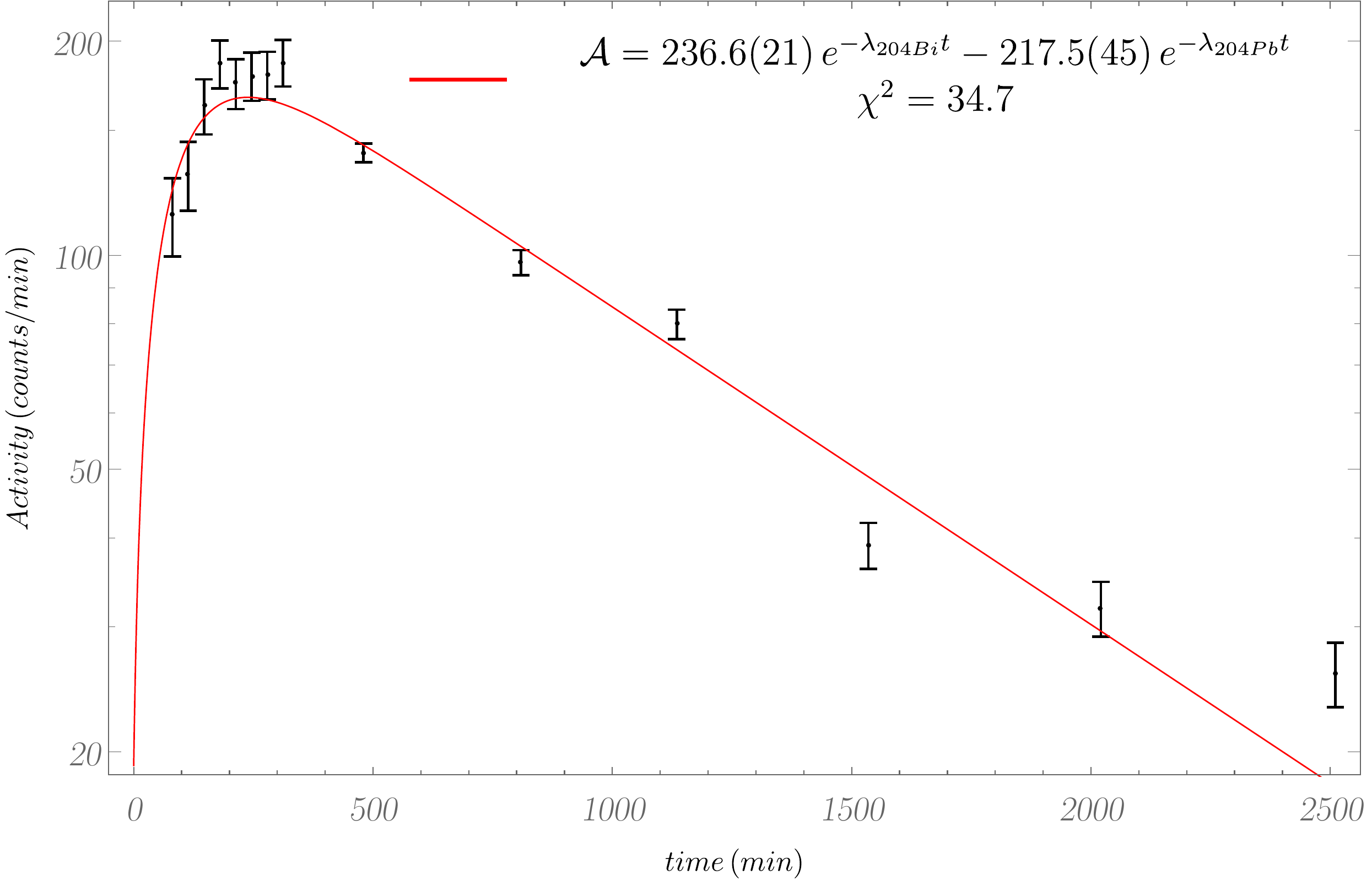}
\caption{Activity decay curve (counts/min)  for $E_\gamma = 532.72 \, \, keV$ from ${}^{204}Bi$ ($1.36 \%$). This fit indicates this transition is possibly fed by the ${}^{204m}Pb$ (66.93 min), although statistically not conclusive.}
\label{fig:Bi204-532}
\end{figure}

Finally we report that some levels of ${}^{204m}Pb$ are probably fed by the decay of ${}^{204}Bi$. Since we did not performed coincidence measurement, we cannot have a conclusive result. ${}^{204}Bi$ EC decay emits two 911 keV $\gamma$'s: 911.96 keV ($11.22 \%$), and 911.74 keV ($13.6 \%$). The 911.74 keV is emitted by the transition of the $2185.88 \mapsto 1274.13 \, \, keV$ ($9^- \mapsto 4^+$). The 2185.88 keV level is a isomeric one with a half-life of 66.93 min. Although not report in nuclear databases, this $\gamma$ represents a delayed one and it is important to consider in astrophysical models. When we follow the decay of ${}^{204}Bi$ by the 911 keV peak, we observe exactly this behavior, as noted in Figure \ref{fig:Bi204-911}. It is emitted by this transition and by another than feed the fundamental level of ${}^{204}Pb$ directly: $E_\gamma = 911.96 \, \, keV$ ($3170.37 \mapsto 2258.15 \, \, keV$; $5^- \mapsto 5^-$). We note a similar behavior for the 532.72 keV $\gamma$, although it is not expected, Figure \ref{fig:Bi204-532}. If it only the gamma emitted by the $3638.05 \mapsto 3105.29 \, \, keV $ $\gamma$-transition, it should not have a delay of $\sim 60 \, min$, as observed. Like in the 911 keV, it could have another transition from 2185.85 keV level (66.93 min). Since our measurements are in the statistical limit of compatibility for a delayed gamma emission, we recommend new measurements to establish the levels fed by ${}^{204}Bi$.

\begin{table}[ht]
	\centering
	\setlength{\tabcolsep}{2pt}
	\caption{Cross-section values $(mb)$ for proton energy in the lab frame for the reactions $ {}^{206}{Pb}(p,xn){}^{204}{Bi}$.
	}
	\label{tab:xs-pb-206}
	\begin{tabular}{|c|c|}
		\hline
		Energy  (MeV) & \({}^{204}{Bi} \) \\ \hline
	30.0 & 631 (7) \\ 
	29.0 & 1215 (42)  \\ 
	28.0 & 838 (10) \\ 
	27.0 & \( \sim \)516 (6) \\ 
	26.0 & \( \sim \)229.9 (29) \\ 
	25.0 & \( \sim \)32.0 (4) \\ 
	24.0 & - \\ 
	23.0 & - \\ 
	22.0 & - \\ 
	21.0 & - \\ 
	20.0 & - 
	\\ \hline
\end{tabular}
\end{table}

\begin{table}[ht]
	\centering
	\setlength{\tabcolsep}{2pt}
	\footnotesize
	\caption{Cross-section values $(mb)$ for proton energy in the lab frame for the reactions $ {}^{204}{Pb}(p,xn){}^{201-204}{Bi}$.
	}
	\label{tab:xs-pb-204}
	\begin{tabular}{|c|c|c|c|c|c|}
		\hline
		Energy (MeV) & ${}^{204}{Bi}$ & ${}^{203}{Bi} $ & ${}^{202}{Bi} $  & ${}^{201}{Bi} $ \\ \hline
	30.0 & - & 323.57 (17) & 349.29 (8) & 24.929 (18) \\ 
	29.0 & - & 635.7 (7) & 765.71 (19) & - \\ 
	28.0 & - & 714.3 (5) & 375.00 (9) & - \\ 
	27.0 & - & 871.4 (5) & 170.04 (4) & - \\ 
	26.0 & - & 942.9 (5) & 36.42 (9) & - \\ 
	25.0 & - & 1007.1 (5) & - & - \\ 
	24.0 & 88.6 (7) & 914.3 (5) & - & - \\ 
	23.0 & 96.43 (7) & 878.6 (5) & - & - \\ 
	22.0 & 127.86 (9) & 785.7 (4) & - & - \\ 
	21.0 & 177.86 (13) & 607.1 (4) & - & - \\ 
	20.0 & 200.71 (15) & 535.7 (3) & - & -
	\\ \hline
\end{tabular}
\end{table}

\section{Discussion}

Previous measurements  of ${}^{nat}Pb(p,xn){}^{201-207}Bi$ cross-sections at energies $< 40 \, MeV$ by \citet{Ditroi_2014,Kuhnhenn_2001,Lagunas-Solar_1987,Bell_1956} have discrepancies among each other of about a factor 2. Due the huge uncertainties of these works, they are compatible among each other, but do not allow to verify small variations in theoretical predictions for these cross-sections. By careful gamma spectroscopy  with  half-lives measurements, we were able to obtain high accuracy values for them and test TALYS code 1.9 simulations for $(p,xn)$ reactions and TENDL-19 libraries, \cite{Koning_2019}. We noticed that these theoretical results overestimate $(p,xn)$ cross-sections in general. When we shift TALYS curve by $\sim 4 \, MeV$, they reproduce the main experimental features. Since lead's isotopes are deformed nuclei, wrong estimation of the Coulomb barrier could led to overestimated results too. Also, wrong neutron binding energy estimation could get the  same discrepancies. \citet{Oranj_2017} compared several nuclear models for $(p,xn)$ reactions on lead and possible combinations of the 24 nuclear input parameters to find the sensitivity of the cross sections to each nuclear model, mainly based on optical models potenticals (OMP), nuclear level density (NLD) and $\gamma$-ray strength functions ($\gamma$SFs). At low energies, all models are pretty simular, but we note a shift of a few MEV among them. In general, JLM-Nolte $\alpha$OMP-HCM-GHFB \cite{Nolte_1987} and KD-$\alpha$OMPIII-HFB-RMF \cite{Koning_2003}, whose major alteration stems from the NLD parameters, have the best results. A linear combination of different models would result in a function that fit cross-sections at low and high energies. Our results, allows us to test these models at a low energy range that was never done before.

We did the first measurement of the ${}^{nat}Pb(p,xn){}^{207}Bi$ at energies below 40 MeV. \citet{Titarenko_2005} studied this reaction for energies between 40-2600 MeV. \citet{Schery_1974} did only one measurement of ${}^{208}Pb(p,n){}^{208}Bi$ cross-section with $E_p = 25.8 MeV$. In general, $(p,2n)$ have higher cross-sections values for heavy elements at these energies, but ${}^{206-208}Pb$ contribute to the production of ${}^{208}Bi$.

We also did the first measurement of the ${}^{204}Pb(p,4n){}^{201}Bi$ cross-section at the threshold energy. The production of neutron deficient nucleus in the threshold is an important probe to test nuclear potentials of p-isotopes. This is of fundamental importance in astrophysical models for nucleosynthesis of heavy rp-elements, whose production in stars is still a subject of research. We note a good agreement between this result and TALYS simulations. 

At 30 MeV, our results are depleted in comparison to TALYS simulations, when $(p,4n)$ reactions are allowed to occur too. Since, for ${}^{204}Pb(p,4n){}^{201}Bi$ we got a good agreement without any shift of the theoretical curve, we think that only $(p,1-3n)$ curves are being overestimated by TALYS. ${}^{nat}Pb(p,xn)$ is a sum of all possible curves involving lead isotopes, so, assuming that $(p,4n)$ curve is better estimated by it, ${}^{nat}Pb(p,xn)$ with $E \geq 30 \, MeV$ should have a lower value, as we measured.

The nuclear structure and decay of ${}^{201-205}Bi$ still needs a careful study. The $\gamma$-transitions of them were measured in works of more than 30 years and several transitions were marked as ``probable'', although never studied again, \cite{Kondev2004_205,Kondev2008_206,Richel_1976,Richel_1978,Dzhelepov_1985,Goring_1974,Kondev_2007,Kondev_2005,Zhu_2008}. We highlight here the case of ${}^{204}Pb$. As expected, we observed that the 911.96 keV emitted from the decay of ${}^{204}Bi$ is due the decay of ${}^{204m}Pb$ and should be noted in NNDS. This has important implications in nucleosynthesis, since this $\gamma$ is a delayed one and should be considered in dynamic models. It is also important in high performance liquid chromatography (HPLC) and  time  differential perturbed angular correlation (TDPAC) in macromolecules, \cite{Friedemann_2004}. The work of \citet{Cross_1970} as adopted in \citet{Chiara_2010} (NDS) measured first the $\gamma$'s and levels and then concluded that the 911.7 keV $\gamma$ was compatible with IT level of ${}^{204m}Pb$. It did not discussed that  the intensity of this $\gamma$ was measured immediately after EOB. Without that, it is not possible to have a precise $\gamma$ intensity when mixing $\gamma$'s from ${}^{204}Pb$ and ${}^{204m}Pb$, since their decay constants are different. Our work could not have a definite answer for this question, because we did not perform a coincidence gamma spectroscopy, although the fit of the 911 keV curve half-life did not support this. 
The 532.72 keV decay curve indicates that this level is possibly being fed by the ${}^{204m}Pb$. There are some levels from the ${}^{204m}Pb$ (2185.88 keV, $9^-$) whose level, spin and parity would allow a $\gamma$ of 532 keV. Our result is in the limit of the confidence level, so we have not a strong evidence. Since the ${}^{204}Bi$ EC decay shows  some discrepancies, we recommend a new measurement of the levels structure of this decay.

The new determination of $\gamma$-rays' intensities from ${}^{205,206}Bi$ agrees with previous results of \citet{Manthuruthil1972} and \citet{Hamilton1972}, but it is in general more precise, due the measurements of decay curves, high precision gamma spectroscopy and data analysis, although some discrepancies have been found. In some cases, we considered these previous studies to subtract negligible $\gamma$'s contamination. This iterative approach, although unusual, when comparing with anti-Compton and coincidence approaches should lead in an error of $\ll 5 \%$, depending on the accuracy of these results. This way opens the possibility of high precision spectroscopy and data analysis (HPSDA) in substitution to more complex experimental arrangements, that should always be considered, whenever possible.

The study of $(p,xn)$ reactions in heavy elements, specially in neutron-deficient isotopes with protons of $E_p < 50 \, MeV$ could represent a possible path for the $rp/r$-process, acting together in a dynamic production called $r^2p$-process, as proposed by \citet{guillaumon2020-r2p}. The high energy protons could be produced by bubbles and jets in neutrino-driven winds from neutrons stars mergers and supernovae with $T \geq 3 \times 10^9 \, K$, followed by a ``freeze-out'', \cite{Woosley1992,Cameron2003,Bliss2017,Howard1993,Arcones2011,Perego_2014}. Together with the $\alpha$-process proposed by \citet{Woosley1992}, this could produce all the $rp/r$-elements, including heavy rp-elements and thorium and uranium, whose nucleosynthesis is not still completely understood. Due that, experimental $(p,xn)$ cross sections in heavy elements and neutron-deficient isotopes are of special importance in nucleosynthesis models.

\section{Conclusions}

We were able to determine the cross-sections for ${}^{nat}Pb(p,xn){}^{201-207}Bi$, ${}^{204}Pb(p,xn){}^{201-204}Bi$ and ${}^{206}Pb(p,xn){}^{204}Bi$ reactions in the range $20-30 \, \, MeV$. Previous works have big discrepancies among each other, although statistically compatible due the big uncertainties. Our results increase dramatically the accuracy of these cross-sections, allowing us to compare with TALYS code and TENDL curves. For $(p,\gamma)$ and $(p,1-3n)$ curves, we needed to shift the curves for about 4 MeV, probably to wrong theoretical estimation of neutron binding energy and deformed Coulomb barrier. We did  the first measurement of ${}^{204}Pb(p,xn){}^{201}Bi$ in the threshold and of ${}^{nat}Pb(p,xn){}^{207}Bi$. Reactions with protons in the threshold for neutron deficient isotopes are an important probe of nuclear potentials in the $\beta^+$ region and will affect nucleosynthesis models for the rp-elements.

We also obtained more precise results for $\gamma$ intensities of ${}^{205,206}Bi$. Our results are in good agreement with previous ones.

The nuclear structure of stable lead isotopes need a reevaluation, since there are still possible transitions that were never confirmed. In the case of ${}^{204}Bi$ decay, we report that the $\gamma$ of 911.77 keV feed the ${}^{204m}Pb$ (66.93 min), representing a delayed $\gamma$ and should be update in Nuclear Data Sheets. We also show indicatives that the 532 keV could being fed by the ${}^{204m}Pb$, although a conclusive study should be made.

\bibliography{bib}

\begin{thebibliography}{34}%
\makeatletter
\providecommand \@ifxundefined [1]{%
 \@ifx{#1\undefined}
}%
\providecommand \@ifnum [1]{%
 \ifnum #1\expandafter \@firstoftwo
 \else \expandafter \@secondoftwo
 \fi
}%
\providecommand \@ifx [1]{%
 \ifx #1\expandafter \@firstoftwo
 \else \expandafter \@secondoftwo
 \fi
}%
\providecommand \natexlab [1]{#1}%
\providecommand \enquote  [1]{``#1''}%
\providecommand \bibnamefont  [1]{#1}%
\providecommand \bibfnamefont [1]{#1}%
\providecommand \citenamefont [1]{#1}%
\providecommand \href@noop [0]{\@secondoftwo}%
\providecommand \href [0]{\begingroup \@sanitize@url \@href}%
\providecommand \@href[1]{\@@startlink{#1}\@@href}%
\providecommand \@@href[1]{\endgroup#1\@@endlink}%
\providecommand \@sanitize@url [0]{\catcode `\\12\catcode `\$12\catcode
  `\&12\catcode `\#12\catcode `\^12\catcode `\_12\catcode `\%12\relax}%
\providecommand \@@startlink[1]{}%
\providecommand \@@endlink[0]{}%
\providecommand \url  [0]{\begingroup\@sanitize@url \@url }%
\providecommand \@url [1]{\endgroup\@href {#1}{\urlprefix }}%
\providecommand \urlprefix  [0]{URL }%
\providecommand \Eprint [0]{\href }%
\providecommand \doibase [0]{http://dx.doi.org/}%
\providecommand \selectlanguage [0]{\@gobble}%
\providecommand \bibinfo  [0]{\@secondoftwo}%
\providecommand \bibfield  [0]{\@secondoftwo}%
\providecommand \translation [1]{[#1]}%
\providecommand \BibitemOpen [0]{}%
\providecommand \bibitemStop [0]{}%
\providecommand \bibitemNoStop [0]{.\EOS\space}%
\providecommand \EOS [0]{\spacefactor3000\relax}%
\providecommand \BibitemShut  [1]{\csname bibitem#1\endcsname}%
\let\auto@bib@innerbib\@empty
\bibitem [{\citenamefont {{Guillaumon}}(2019)}]{Guillaumon_2019}%
  \BibitemOpen
  \bibfield  {author} {\bibinfo {author} {\bibfnamefont {P.~V.}\ \bibnamefont
  {{Guillaumon}}},\ }\emph {\bibinfo {title} {Study of reactions
  ${}^{nat}Pb(p,xn){}^{201-207}Bi$ and possible implications for the
  r-process}},\ \href@noop {} {Ph.D. thesis},\ \bibinfo  {school} {Universidade
  de Sao Paulo} (\bibinfo {year} {2019})\BibitemShut {NoStop}%
\bibitem [{\citenamefont {Gouffon}(1987)}]{Idefix}%
  \BibitemOpen
  \bibfield  {author} {\bibinfo {author} {\bibfnamefont {P.}~\bibnamefont
  {Gouffon}},\ }\href@noop {} {\emph {\bibinfo {title} {{Manual do programa
  IDEFIX}}}}\ (\bibinfo  {publisher} {Instituto de Física da Universidade de
  São Paulo, Laboratório do Acelerador Linear},\ \bibinfo {address} {São
  Paulo},\ \bibinfo {year} {1987})\BibitemShut {NoStop}%
\bibitem [{\citenamefont {Manthuruthil}\ \emph {et~al.}(1972)\citenamefont
  {Manthuruthil}, \citenamefont {Camp}, \citenamefont {Ramayya}, \citenamefont
  {Hamilton}, \citenamefont {Pinajian},\ and\ \citenamefont
  {Doornebos}}]{Manthuruthil1972}%
  \BibitemOpen
  \bibfield  {author} {\bibinfo {author} {\bibfnamefont {J.~C.}\ \bibnamefont
  {Manthuruthil}}, \bibinfo {author} {\bibfnamefont {D.~C.}\ \bibnamefont
  {Camp}}, \bibinfo {author} {\bibfnamefont {A.~V.}\ \bibnamefont {Ramayya}},
  \bibinfo {author} {\bibfnamefont {J.~H.}\ \bibnamefont {Hamilton}}, \bibinfo
  {author} {\bibfnamefont {J.~J.}\ \bibnamefont {Pinajian}}, \ and\ \bibinfo
  {author} {\bibfnamefont {J.~W.}\ \bibnamefont {Doornebos}},\ }\href {\doibase
  10.1103/PhysRevC.6.1870} {\bibfield  {journal} {\bibinfo  {journal} {Phys.
  Rev. C}\ }\textbf {\bibinfo {volume} {6}},\ \bibinfo {pages} {1870} (\bibinfo
  {year} {1972})}\BibitemShut {NoStop}%
\bibitem [{\citenamefont {Hamilton}\ \emph {et~al.}(1972)\citenamefont
  {Hamilton}, \citenamefont {Ananthakrishnan}, \citenamefont {Ramayya},
  \citenamefont {LaCasse}, \citenamefont {Camp}, \citenamefont {Pinajian},
  \citenamefont {Kern},\ and\ \citenamefont {Manthuruthil}}]{Hamilton1972}%
  \BibitemOpen
  \bibfield  {author} {\bibinfo {author} {\bibfnamefont {J.~H.}\ \bibnamefont
  {Hamilton}}, \bibinfo {author} {\bibfnamefont {V.}~\bibnamefont
  {Ananthakrishnan}}, \bibinfo {author} {\bibfnamefont {A.~V.}\ \bibnamefont
  {Ramayya}}, \bibinfo {author} {\bibfnamefont {W.~M.}\ \bibnamefont
  {LaCasse}}, \bibinfo {author} {\bibfnamefont {D.~C.}\ \bibnamefont {Camp}},
  \bibinfo {author} {\bibfnamefont {J.~J.}\ \bibnamefont {Pinajian}}, \bibinfo
  {author} {\bibfnamefont {L.~H.}\ \bibnamefont {Kern}}, \ and\ \bibinfo
  {author} {\bibfnamefont {J.~C.}\ \bibnamefont {Manthuruthil}},\ }\href
  {\doibase 10.1103/PhysRevC.6.1265} {\bibfield  {journal} {\bibinfo  {journal}
  {Phys. Rev. C}\ }\textbf {\bibinfo {volume} {6}},\ \bibinfo {pages} {1265}
  (\bibinfo {year} {1972})}\BibitemShut {NoStop}%
\bibitem [{\citenamefont {Kondev}(2004)}]{Kondev2004_205}%
  \BibitemOpen
  \bibfield  {author} {\bibinfo {author} {\bibfnamefont {F.}~\bibnamefont
  {Kondev}},\ }\href {\doibase https://doi.org/10.1016/j.nds.2004.03.001}
  {\bibfield  {journal} {\bibinfo  {journal} {Nuclear Data Sheets}\ }\textbf
  {\bibinfo {volume} {101}},\ \bibinfo {pages} {521 } (\bibinfo {year}
  {2004})}\BibitemShut {NoStop}%
\bibitem [{\citenamefont {Kondev}(2008)}]{Kondev2008_206}%
  \BibitemOpen
  \bibfield  {author} {\bibinfo {author} {\bibfnamefont {F.}~\bibnamefont
  {Kondev}},\ }\href {\doibase https://doi.org/10.1016/j.nds.2008.05.002}
  {\bibfield  {journal} {\bibinfo  {journal} {Nuclear Data Sheets}\ }\textbf
  {\bibinfo {volume} {109}},\ \bibinfo {pages} {1527 } (\bibinfo {year}
  {2008})}\BibitemShut {NoStop}%
\bibitem [{\citenamefont {Koning}\ and\ \citenamefont
  {Rochman}(2012)}]{Koning_2012}%
  \BibitemOpen
  \bibfield  {author} {\bibinfo {author} {\bibfnamefont {A.}~\bibnamefont
  {Koning}}\ and\ \bibinfo {author} {\bibfnamefont {D.}~\bibnamefont
  {Rochman}},\ }\href {\doibase https://doi.org/10.1016/j.nds.2012.11.002}
  {\bibfield  {journal} {\bibinfo  {journal} {Nuclear Data Sheets}\ }\textbf
  {\bibinfo {volume} {113}},\ \bibinfo {pages} {2841 } (\bibinfo {year}
  {2012})},\ \bibinfo {note} {special Issue on Nuclear Reaction
  Data}\BibitemShut {NoStop}%
\bibitem [{\citenamefont {Kuhnhenn}\ \emph {et~al.}(2001)\citenamefont
  {Kuhnhenn}, \citenamefont {Herpers}, \citenamefont {Glasser}, \citenamefont
  {Michel}, \citenamefont {Kubik},\ and\ \citenamefont
  {Suter}}]{Kuhnhenn_2001}%
  \BibitemOpen
  \bibfield  {author} {\bibinfo {author} {\bibfnamefont {J.}~\bibnamefont
  {Kuhnhenn}}, \bibinfo {author} {\bibfnamefont {U.}~\bibnamefont {Herpers}},
  \bibinfo {author} {\bibfnamefont {W.}~\bibnamefont {Glasser}}, \bibinfo
  {author} {\bibfnamefont {R.}~\bibnamefont {Michel}}, \bibinfo {author}
  {\bibfnamefont {P.~W.}\ \bibnamefont {Kubik}}, \ and\ \bibinfo {author}
  {\bibfnamefont {M.}~\bibnamefont {Suter}},\ }\href
  {http://inis.iaea.org/search/search.aspx?orig_q=RN:33011657} {\bibfield
  {journal} {\bibinfo  {journal} {Radiochimica Acta}\ }\textbf {\bibinfo
  {volume} {89}},\ \bibinfo {pages} {697} (\bibinfo {year} {2001})}\BibitemShut
  {NoStop}%
\bibitem [{\citenamefont {Ditrói}\ \emph {et~al.}(2014)\citenamefont
  {Ditrói}, \citenamefont {Tárkányi}, \citenamefont {Takács},\ and\
  \citenamefont {Hermanne}}]{Ditroi_2014}%
  \BibitemOpen
  \bibfield  {author} {\bibinfo {author} {\bibfnamefont {F.}~\bibnamefont
  {Ditrói}}, \bibinfo {author} {\bibfnamefont {F.}~\bibnamefont {Tárkányi}},
  \bibinfo {author} {\bibfnamefont {S.}~\bibnamefont {Takács}}, \ and\
  \bibinfo {author} {\bibfnamefont {A.}~\bibnamefont {Hermanne}},\ }\href
  {\doibase https://doi.org/10.1016/j.apradiso.2014.04.006} {\bibfield
  {journal} {\bibinfo  {journal} {Applied Radiation and Isotopes}\ }\textbf
  {\bibinfo {volume} {90}},\ \bibinfo {pages} {208 } (\bibinfo {year}
  {2014})}\BibitemShut {NoStop}%
\bibitem [{\citenamefont {Lagunas-Solar}\ \emph {et~al.}(1987)\citenamefont
  {Lagunas-Solar}, \citenamefont {Carvacho}, \citenamefont {Nagahara},
  \citenamefont {Mishra},\ and\ \citenamefont {Parks}}]{Lagunas-Solar_1987}%
  \BibitemOpen
  \bibfield  {author} {\bibinfo {author} {\bibfnamefont {M.~C.}\ \bibnamefont
  {Lagunas-Solar}}, \bibinfo {author} {\bibfnamefont {O.~F.}\ \bibnamefont
  {Carvacho}}, \bibinfo {author} {\bibfnamefont {L.}~\bibnamefont {Nagahara}},
  \bibinfo {author} {\bibfnamefont {A.}~\bibnamefont {Mishra}}, \ and\ \bibinfo
  {author} {\bibfnamefont {N.~J.}\ \bibnamefont {Parks}},\ }\href {\doibase
  10.1016/0883-2889(87)90008-6} {\bibfield  {journal} {\bibinfo  {journal}
  {Int. J. Radiat. Appl. Instrum. Part A}\ }\textbf {\bibinfo {volume} {38}},\
  \bibinfo {pages} {129} (\bibinfo {year} {1987})}\BibitemShut {NoStop}%
\bibitem [{\citenamefont {Bell}\ and\ \citenamefont
  {Skarsgard}(1956)}]{Bell_1956}%
  \BibitemOpen
  \bibfield  {author} {\bibinfo {author} {\bibfnamefont {R.~E.}\ \bibnamefont
  {Bell}}\ and\ \bibinfo {author} {\bibfnamefont {H.~M.}\ \bibnamefont
  {Skarsgard}},\ }\href {\doibase 10.1139/p56-086} {\bibfield  {journal}
  {\bibinfo  {journal} {Canadian Journal of Physics}\ }\textbf {\bibinfo
  {volume} {34}},\ \bibinfo {pages} {745} (\bibinfo {year} {1956})},\ \Eprint
  {http://arxiv.org/abs/https://doi.org/10.1139/p56-086}
  {https://doi.org/10.1139/p56-086} \BibitemShut {NoStop}%
\bibitem [{\citenamefont {Koning}\ \emph {et~al.}(2019)\citenamefont {Koning},
  \citenamefont {Rochman}, \citenamefont {Sublet}, \citenamefont {Dzysiuk},
  \citenamefont {Fleming},\ and\ \citenamefont {{van der
  Marck}}}]{Koning_2019}%
  \BibitemOpen
  \bibfield  {author} {\bibinfo {author} {\bibfnamefont {A.}~\bibnamefont
  {Koning}}, \bibinfo {author} {\bibfnamefont {D.}~\bibnamefont {Rochman}},
  \bibinfo {author} {\bibfnamefont {J.-C.}\ \bibnamefont {Sublet}}, \bibinfo
  {author} {\bibfnamefont {N.}~\bibnamefont {Dzysiuk}}, \bibinfo {author}
  {\bibfnamefont {M.}~\bibnamefont {Fleming}}, \ and\ \bibinfo {author}
  {\bibfnamefont {S.}~\bibnamefont {{van der Marck}}},\ }\href {\doibase
  https://doi.org/10.1016/j.nds.2019.01.002} {\bibfield  {journal} {\bibinfo
  {journal} {Nuclear Data Sheets}\ }\textbf {\bibinfo {volume} {155}},\
  \bibinfo {pages} {1 } (\bibinfo {year} {2019})},\ \bibinfo {note} {special
  Issue on Nuclear Reaction Data}\BibitemShut {NoStop}%
\bibitem [{\citenamefont {{Mokhtari Oranj}}\ \emph {et~al.}(2017)\citenamefont
  {{Mokhtari Oranj}}, \citenamefont {{Jung}}, \citenamefont {{Bakhtiari}},
  \citenamefont {{Lee}},\ and\ \citenamefont {{Lee}}}]{Oranj_2017}%
  \BibitemOpen
  \bibfield  {author} {\bibinfo {author} {\bibfnamefont {L.}~\bibnamefont
  {{Mokhtari Oranj}}}, \bibinfo {author} {\bibfnamefont {N.~S.}\ \bibnamefont
  {{Jung}}}, \bibinfo {author} {\bibfnamefont {M.}~\bibnamefont {{Bakhtiari}}},
  \bibinfo {author} {\bibfnamefont {A.}~\bibnamefont {{Lee}}}, \ and\ \bibinfo
  {author} {\bibfnamefont {H.~S.}\ \bibnamefont {{Lee}}},\ }\href {\doibase
  10.1103/PhysRevC.95.044609} {\bibfield  {journal} {\bibinfo  {journal}
  {Physical Review C}\ }\textbf {\bibinfo {volume} {95}},\ \bibinfo {eid}
  {044609} (\bibinfo {year} {2017})}\BibitemShut {NoStop}%
\bibitem [{\citenamefont {Nolte}\ \emph {et~al.}(1987)\citenamefont {Nolte},
  \citenamefont {Machner},\ and\ \citenamefont {Bojowald}}]{Nolte_1987}%
  \BibitemOpen
  \bibfield  {author} {\bibinfo {author} {\bibfnamefont {M.}~\bibnamefont
  {Nolte}}, \bibinfo {author} {\bibfnamefont {H.}~\bibnamefont {Machner}}, \
  and\ \bibinfo {author} {\bibfnamefont {J.}~\bibnamefont {Bojowald}},\ }\href
  {\doibase 10.1103/PhysRevC.36.1312} {\bibfield  {journal} {\bibinfo
  {journal} {Phys. Rev. C}\ }\textbf {\bibinfo {volume} {36}},\ \bibinfo
  {pages} {1312} (\bibinfo {year} {1987})}\BibitemShut {NoStop}%
\bibitem [{\citenamefont {Koning}\ and\ \citenamefont
  {Delaroche}(2003)}]{Koning_2003}%
  \BibitemOpen
  \bibfield  {author} {\bibinfo {author} {\bibfnamefont {A.}~\bibnamefont
  {Koning}}\ and\ \bibinfo {author} {\bibfnamefont {J.}~\bibnamefont
  {Delaroche}},\ }\href {\doibase
  https://doi.org/10.1016/S0375-9474(02)01321-0} {\bibfield  {journal}
  {\bibinfo  {journal} {Nuclear Physics A}\ }\textbf {\bibinfo {volume}
  {713}},\ \bibinfo {pages} {231 } (\bibinfo {year} {2003})}\BibitemShut
  {NoStop}%
\bibitem [{\citenamefont {Titarenko}\ \emph {et~al.}(2005)\citenamefont
  {Titarenko}, \citenamefont {Batyaev}, \citenamefont {Zhivun}, \citenamefont
  {Mulambetov}, \citenamefont {Mulambetova}, \citenamefont {Zaitsev},
  \citenamefont {Mashnik},\ and\ \citenamefont {Prael}}]{Titarenko_2005}%
  \BibitemOpen
  \bibfield  {author} {\bibinfo {author} {\bibfnamefont {Y.~E.}\ \bibnamefont
  {Titarenko}}, \bibinfo {author} {\bibfnamefont {V.~F.}\ \bibnamefont
  {Batyaev}}, \bibinfo {author} {\bibfnamefont {V.~M.}\ \bibnamefont {Zhivun}},
  \bibinfo {author} {\bibfnamefont {R.~D.}\ \bibnamefont {Mulambetov}},
  \bibinfo {author} {\bibfnamefont {S.~V.}\ \bibnamefont {Mulambetova}},
  \bibinfo {author} {\bibfnamefont {S.~L.}\ \bibnamefont {Zaitsev}}, \bibinfo
  {author} {\bibfnamefont {S.~G.}\ \bibnamefont {Mashnik}}, \ and\ \bibinfo
  {author} {\bibfnamefont {R.~E.}\ \bibnamefont {Prael}},\ }\href {\doibase
  10.1063/1.1945192} {\bibfield  {journal} {\bibinfo  {journal} {AIP Conference
  Proceedings}\ }\textbf {\bibinfo {volume} {769}},\ \bibinfo {pages} {1070}
  (\bibinfo {year} {2005})},\ \Eprint
  {http://arxiv.org/abs/https://aip.scitation.org/doi/pdf/10.1063/1.1945192}
  {https://aip.scitation.org/doi/pdf/10.1063/1.1945192} \BibitemShut {NoStop}%
\bibitem [{\citenamefont {Schery}\ \emph {et~al.}(1974)\citenamefont {Schery},
  \citenamefont {Lind}, \citenamefont {Fielding},\ and\ \citenamefont
  {Zafiratos}}]{Schery_1974}%
  \BibitemOpen
  \bibfield  {author} {\bibinfo {author} {\bibfnamefont {S.}~\bibnamefont
  {Schery}}, \bibinfo {author} {\bibfnamefont {D.}~\bibnamefont {Lind}},
  \bibinfo {author} {\bibfnamefont {H.}~\bibnamefont {Fielding}}, \ and\
  \bibinfo {author} {\bibfnamefont {C.}~\bibnamefont {Zafiratos}},\ }\href
  {\doibase https://doi.org/10.1016/0375-9474(74)90382-0} {\bibfield  {journal}
  {\bibinfo  {journal} {Nuclear Physics A}\ }\textbf {\bibinfo {volume}
  {234}},\ \bibinfo {pages} {109 } (\bibinfo {year} {1974})}\BibitemShut
  {NoStop}%
\bibitem [{\citenamefont {Richel}\ \emph {et~al.}(1976)\citenamefont {Richel},
  \citenamefont {Albouy}, \citenamefont {Auger}, \citenamefont {David},
  \citenamefont {Lagrange}, \citenamefont {Pautrat}, \citenamefont {Roulet},
  \citenamefont {Sergolle},\ and\ \citenamefont {Vanhorenbeeck}}]{Richel_1976}%
  \BibitemOpen
  \bibfield  {author} {\bibinfo {author} {\bibfnamefont {H.}~\bibnamefont
  {Richel}}, \bibinfo {author} {\bibfnamefont {G.}~\bibnamefont {Albouy}},
  \bibinfo {author} {\bibfnamefont {G.}~\bibnamefont {Auger}}, \bibinfo
  {author} {\bibfnamefont {J.}~\bibnamefont {David}}, \bibinfo {author}
  {\bibfnamefont {J.}~\bibnamefont {Lagrange}}, \bibinfo {author}
  {\bibfnamefont {M.}~\bibnamefont {Pautrat}}, \bibinfo {author} {\bibfnamefont
  {C.}~\bibnamefont {Roulet}}, \bibinfo {author} {\bibfnamefont
  {H.}~\bibnamefont {Sergolle}}, \ and\ \bibinfo {author} {\bibfnamefont
  {J.}~\bibnamefont {Vanhorenbeeck}},\ }\href {\doibase
  https://doi.org/10.1016/0375-9474(76)90658-8} {\bibfield  {journal} {\bibinfo
   {journal} {Nuclear Physics A}\ }\textbf {\bibinfo {volume} {267}},\ \bibinfo
  {pages} {253 } (\bibinfo {year} {1976})}\BibitemShut {NoStop}%
\bibitem [{\citenamefont {Richel}\ \emph {et~al.}(1978)\citenamefont {Richel},
  \citenamefont {Albouy}, \citenamefont {Auger}, \citenamefont {Hanappe},
  \citenamefont {Lagrange}, \citenamefont {Pautrat}, \citenamefont {Roulet},
  \citenamefont {Sergolle},\ and\ \citenamefont {Vanhorenbeeck}}]{Richel_1978}%
  \BibitemOpen
  \bibfield  {author} {\bibinfo {author} {\bibfnamefont {H.}~\bibnamefont
  {Richel}}, \bibinfo {author} {\bibfnamefont {G.}~\bibnamefont {Albouy}},
  \bibinfo {author} {\bibfnamefont {G.}~\bibnamefont {Auger}}, \bibinfo
  {author} {\bibfnamefont {F.}~\bibnamefont {Hanappe}}, \bibinfo {author}
  {\bibfnamefont {J.}~\bibnamefont {Lagrange}}, \bibinfo {author}
  {\bibfnamefont {M.}~\bibnamefont {Pautrat}}, \bibinfo {author} {\bibfnamefont
  {C.}~\bibnamefont {Roulet}}, \bibinfo {author} {\bibfnamefont
  {H.}~\bibnamefont {Sergolle}}, \ and\ \bibinfo {author} {\bibfnamefont
  {J.}~\bibnamefont {Vanhorenbeeck}},\ }\href {\doibase
  https://doi.org/10.1016/0375-9474(78)90372-X} {\bibfield  {journal} {\bibinfo
   {journal} {Nuclear Physics A}\ }\textbf {\bibinfo {volume} {303}},\ \bibinfo
  {pages} {483 } (\bibinfo {year} {1978})}\BibitemShut {NoStop}%
\bibitem [{\citenamefont {{Dzhelepov}}\ \emph {et~al.}(1985)\citenamefont
  {{Dzhelepov}}, \citenamefont {{Kuznetsova}}, \citenamefont {{Popova}},\ and\
  \citenamefont {{Prikhodtseva}}}]{Dzhelepov_1985}%
  \BibitemOpen
  \bibfield  {author} {\bibinfo {author} {\bibfnamefont {B.~S.}\ \bibnamefont
  {{Dzhelepov}}}, \bibinfo {author} {\bibfnamefont {M.~Y.}\ \bibnamefont
  {{Kuznetsova}}}, \bibinfo {author} {\bibfnamefont {T.~I.}\ \bibnamefont
  {{Popova}}}, \ and\ \bibinfo {author} {\bibfnamefont {V.~P.}\ \bibnamefont
  {{Prikhodtseva}}},\ }\href@noop {} {\bibfield  {journal} {\bibinfo  {journal}
  {Izv.Akad.Nauk}\ }\textbf {\bibinfo {volume} {SSSR}},\ \bibinfo {pages}
  {Ser.Fiz. 49, 2082} (\bibinfo {year} {1985})}\BibitemShut {NoStop}%
\bibitem [{\citenamefont {{Goring}}\ and\ \citenamefont
  {{Hanser}}(1974)}]{Goring_1974}%
  \BibitemOpen
  \bibfield  {author} {\bibinfo {author} {\bibfnamefont {S.}~\bibnamefont
  {{Goring}}}\ and\ \bibinfo {author} {\bibfnamefont {A.}~\bibnamefont
  {{Hanser}}},\ }\href@noop {} {\bibfield  {journal} {\bibinfo  {journal}
  {Z.Phys.}\ }\textbf {\bibinfo {volume} {271}},\ \bibinfo {pages} {183}
  (\bibinfo {year} {1974})}\BibitemShut {NoStop}%
\bibitem [{\citenamefont {Kondev}(2007)}]{Kondev_2007}%
  \BibitemOpen
  \bibfield  {author} {\bibinfo {author} {\bibfnamefont {F.}~\bibnamefont
  {Kondev}},\ }\href {\doibase https://doi.org/10.1016/j.nds.2007.01.004}
  {\bibfield  {journal} {\bibinfo  {journal} {Nuclear Data Sheets}\ }\textbf
  {\bibinfo {volume} {108}},\ \bibinfo {pages} {365 } (\bibinfo {year}
  {2007})}\BibitemShut {NoStop}%
\bibitem [{\citenamefont {Kondev}(2005)}]{Kondev_2005}%
  \BibitemOpen
  \bibfield  {author} {\bibinfo {author} {\bibfnamefont {F.}~\bibnamefont
  {Kondev}},\ }\href {\doibase https://doi.org/10.1016/j.nds.2005.05.001}
  {\bibfield  {journal} {\bibinfo  {journal} {Nuclear Data Sheets}\ }\textbf
  {\bibinfo {volume} {105}},\ \bibinfo {pages} {1 } (\bibinfo {year}
  {2005})}\BibitemShut {NoStop}%
\bibitem [{\citenamefont {Zhu}\ and\ \citenamefont {Kondev}(2008)}]{Zhu_2008}%
  \BibitemOpen
  \bibfield  {author} {\bibinfo {author} {\bibfnamefont {S.}~\bibnamefont
  {Zhu}}\ and\ \bibinfo {author} {\bibfnamefont {F.}~\bibnamefont {Kondev}},\
  }\href {\doibase https://doi.org/10.1016/j.nds.2008.02.002} {\bibfield
  {journal} {\bibinfo  {journal} {Nuclear Data Sheets}\ }\textbf {\bibinfo
  {volume} {109}},\ \bibinfo {pages} {699 } (\bibinfo {year}
  {2008})}\BibitemShut {NoStop}%
\bibitem [{\citenamefont {Friedemann}\ \emph {et~al.}(2004)\citenamefont
  {Friedemann}, \citenamefont {Heinrich}, \citenamefont {Haas},\ and\
  \citenamefont {Tröger}}]{Friedemann_2004}%
  \BibitemOpen
  \bibfield  {author} {\bibinfo {author} {\bibfnamefont {S.}~\bibnamefont
  {Friedemann}}, \bibinfo {author} {\bibfnamefont {F.}~\bibnamefont
  {Heinrich}}, \bibinfo {author} {\bibfnamefont {H.}~\bibnamefont {Haas}}, \
  and\ \bibinfo {author} {\bibfnamefont {W.}~\bibnamefont {Tröger}},\ }\href
  {\doibase 10.1007/s10751-005-9121-4} {\bibfield  {journal} {\bibinfo
  {journal} {Hyperfine Interactions}\ }\textbf {\bibinfo {volume} {159}},\
  \bibinfo {pages} {313} (\bibinfo {year} {2004})}\BibitemShut {NoStop}%
\bibitem [{\citenamefont {{Cross}}(1970)}]{Cross_1970}%
  \BibitemOpen
  \bibfield  {author} {\bibinfo {author} {\bibfnamefont {J.~B.}\ \bibnamefont
  {{Cross}}},\ }\emph {\bibinfo {title} {The Electron Capture Decay Schemes of
  Bi$^{203}$ and Bi$^{204}$}},\ \href@noop {} {Ph.D. thesis},\ \bibinfo
  {school} {Michigan State Univ.} (\bibinfo {year} {1970})\BibitemShut
  {NoStop}%
\bibitem [{\citenamefont {Chiara}\ and\ \citenamefont
  {Kondev}(2010)}]{Chiara_2010}%
  \BibitemOpen
  \bibfield  {author} {\bibinfo {author} {\bibfnamefont {C.}~\bibnamefont
  {Chiara}}\ and\ \bibinfo {author} {\bibfnamefont {F.}~\bibnamefont
  {Kondev}},\ }\href {\doibase 10.1016/j.nds.2009.12.002} {\bibfield  {journal}
  {\bibinfo  {journal} {Nucl. Data Sheets}\ }\textbf {\bibinfo {volume}
  {111}},\ \bibinfo {pages} {141} (\bibinfo {year} {2010})}\BibitemShut
  {NoStop}%
\bibitem [{\citenamefont {Guillaumon}\ and\ \citenamefont
  {Goldman}(2020)}]{guillaumon2020-r2p}%
  \BibitemOpen
  \bibfield  {author} {\bibinfo {author} {\bibfnamefont {P.~V.}\ \bibnamefont
  {Guillaumon}}\ and\ \bibinfo {author} {\bibfnamefont {I.~D.}\ \bibnamefont
  {Goldman}},\ }\href@noop {} {\enquote {\bibinfo {title} {The importance of
  charged particle reactions in the r-process on supernovae and neutron
  stars},}\ } (\bibinfo {year} {2020}),\ \Eprint
  {http://arxiv.org/abs/2009.01814} {arXiv:2009.01814 [nucl-th]} \BibitemShut
  {NoStop}%
\bibitem [{\citenamefont {{Woosley}}\ and\ \citenamefont
  {{Hoffman}}(1992)}]{Woosley1992}%
  \BibitemOpen
  \bibfield  {author} {\bibinfo {author} {\bibfnamefont {S.~E.}\ \bibnamefont
  {{Woosley}}}\ and\ \bibinfo {author} {\bibfnamefont {R.~D.}\ \bibnamefont
  {{Hoffman}}},\ }\href {\doibase 10.1086/171644} {\bibfield  {journal}
  {\bibinfo  {journal} {\apj}\ }\textbf {\bibinfo {volume} {395}},\ \bibinfo
  {pages} {202} (\bibinfo {year} {1992})}\BibitemShut {NoStop}%
\bibitem [{\citenamefont {{Cameron}}(2003)}]{Cameron2003}%
  \BibitemOpen
  \bibfield  {author} {\bibinfo {author} {\bibfnamefont {A.~G.~W.}\
  \bibnamefont {{Cameron}}},\ }\href {\doibase 10.1086/368110} {\bibfield
  {journal} {\bibinfo  {journal} {The Astrophysical Journal}\ }\textbf
  {\bibinfo {volume} {587}},\ \bibinfo {pages} {327} (\bibinfo {year}
  {2003})}\BibitemShut {NoStop}%
\bibitem [{\citenamefont {{Bliss}}\ \emph {et~al.}(2017)\citenamefont
  {{Bliss}}, \citenamefont {{Arcones}}, \citenamefont {{Montes}},\ and\
  \citenamefont {{Pereira}}}]{Bliss2017}%
  \BibitemOpen
  \bibfield  {author} {\bibinfo {author} {\bibfnamefont {J.}~\bibnamefont
  {{Bliss}}}, \bibinfo {author} {\bibfnamefont {A.}~\bibnamefont {{Arcones}}},
  \bibinfo {author} {\bibfnamefont {F.}~\bibnamefont {{Montes}}}, \ and\
  \bibinfo {author} {\bibfnamefont {J.}~\bibnamefont {{Pereira}}},\ }\href
  {\doibase 10.1088/1361-6471/aa63bd} {\bibfield  {journal} {\bibinfo
  {journal} {Journal of Physics G Nuclear Physics}\ }\textbf {\bibinfo {volume}
  {44}},\ \bibinfo {eid} {054003} (\bibinfo {year} {2017})},\ \Eprint
  {http://arxiv.org/abs/1612.02435} {arXiv:1612.02435 [astro-ph.SR]}
  \BibitemShut {NoStop}%
\bibitem [{\citenamefont {{Howard}}\ \emph {et~al.}(1993)\citenamefont
  {{Howard}}, \citenamefont {{Goriely}}, \citenamefont {{Rayet}},\ and\
  \citenamefont {{Arnould}}}]{Howard1993}%
  \BibitemOpen
  \bibfield  {author} {\bibinfo {author} {\bibfnamefont {W.~M.}\ \bibnamefont
  {{Howard}}}, \bibinfo {author} {\bibfnamefont {S.}~\bibnamefont {{Goriely}}},
  \bibinfo {author} {\bibfnamefont {M.}~\bibnamefont {{Rayet}}}, \ and\
  \bibinfo {author} {\bibfnamefont {M.}~\bibnamefont {{Arnould}}},\ }\href
  {\doibase 10.1086/173350} {\bibfield  {journal} {\bibinfo  {journal} {\apj}\
  }\textbf {\bibinfo {volume} {417}},\ \bibinfo {pages} {713} (\bibinfo {year}
  {1993})}\BibitemShut {NoStop}%
\bibitem [{\citenamefont {Arcones}\ and\ \citenamefont
  {Mart\'{\i}nez-Pinedo}(2011)}]{Arcones2011}%
  \BibitemOpen
  \bibfield  {author} {\bibinfo {author} {\bibfnamefont {A.}~\bibnamefont
  {Arcones}}\ and\ \bibinfo {author} {\bibfnamefont {G.}~\bibnamefont
  {Mart\'{\i}nez-Pinedo}},\ }\href {\doibase 10.1103/PhysRevC.83.045809}
  {\bibfield  {journal} {\bibinfo  {journal} {Phys. Rev. C}\ }\textbf {\bibinfo
  {volume} {83}},\ \bibinfo {pages} {045809} (\bibinfo {year}
  {2011})}\BibitemShut {NoStop}%
\bibitem [{\citenamefont {Perego}\ \emph {et~al.}(2014)\citenamefont {Perego},
  \citenamefont {Rosswog}, \citenamefont {Cabezón}, \citenamefont {Korobkin},
  \citenamefont {Käppeli}, \citenamefont {Arcones},\ and\ \citenamefont
  {Liebendörfer}}]{Perego_2014}%
  \BibitemOpen
  \bibfield  {author} {\bibinfo {author} {\bibfnamefont {A.}~\bibnamefont
  {Perego}}, \bibinfo {author} {\bibfnamefont {S.}~\bibnamefont {Rosswog}},
  \bibinfo {author} {\bibfnamefont {R.~M.}\ \bibnamefont {Cabezón}}, \bibinfo
  {author} {\bibfnamefont {O.}~\bibnamefont {Korobkin}}, \bibinfo {author}
  {\bibfnamefont {R.}~\bibnamefont {Käppeli}}, \bibinfo {author}
  {\bibfnamefont {A.}~\bibnamefont {Arcones}}, \ and\ \bibinfo {author}
  {\bibfnamefont {M.}~\bibnamefont {Liebendörfer}},\ }\href {\doibase
  10.1093/mnras/stu1352} {\bibfield  {journal} {\bibinfo  {journal} {Monthly
  Notices of the Royal Astronomical Society}\ }\textbf {\bibinfo {volume}
  {443}},\ \bibinfo {pages} {3134} (\bibinfo {year} {2014})},\ \Eprint
  {http://arxiv.org/abs/https://academic.oup.com/mnras/article-pdf/443/4/3134/6276168/stu1352.pdf}
  {https://academic.oup.com/mnras/article-pdf/443/4/3134/6276168/stu1352.pdf}
  \BibitemShut {NoStop}%
\end{thebibliography}%

\end{document}